\documentclass[letterpaper,11pt]{article}
\usepackage{latexsym}
\usepackage{amsfonts}
\usepackage{amsmath}
\usepackage{listings}
\usepackage{boxedminipage}
\usepackage{url}
\usepackage{graphicx}
\usepackage{subfigure}
\usepackage{color}
\usepackage{multirow}

\newtheorem{theorem}{Theorem}

\newtheorem{definition}{Definition}

\newtheorem{example}{Example}

\newcommand{\negl}{\mathtt{negl}}

\newcommand{\comment}[1]{}              
\newcommand{\from}{\leftarrow}
\newcommand{\aoe}{\ensuremath{\mathsf{AOE}}}
\newcommand{\aoeName}{Amortized Orthogonality Encryption}
\newcommand{\sssName}{Secure Selective Stream}
\newcommand{\sss}{\ensuremath{\mathsf{SSS}}}
\newcommand{\sym}{\ensuremath{\mathsf{SYM}}}

\newcommand{\inst}{\mathcal{I}}  
\newcommand{\I}{\inst}           

\newcommand{\poly}{{\mathtt{poly}}}

\newcommand{\DS}{{\sf DS}}        
\newcommand{\DO}{{\sf DO}}        
\newcommand{\QP}{{\sf QP}}        
\newcommand{\QS}{{\sf QS}}        
\newcommand{\CH}{{\sf Ch}}        
\newcommand{\ADV}{{\cal A}}       
\newcommand{\ADVA}{{\ADV}}        
\newcommand{\ADVB}{{\cal B}}      

\newcommand{\row}{\mathtt{Row}}

\newcommand{\pol}{\mathtt{Pol}}

\newcommand{\accReq}{{\mathsf{AccReq}}}
\newcommand{\stream}{{\mathsf{Stream}}}
\newcommand{\estream}{{\mathsf{e}\stream}}
\newcommand{\val}{\mathtt{val}}

\newcommand{\X}{{\mathcal X}}

\newcommand{\V}{{\mathcal V}}
\newcommand{\ZZ}{{\mathcal Z}}
\newcommand{\M}{{\mathcal M}}
\newcommand{\HH}{{\mathcal H}}

\newcommand{\C}{{\mathcal C}}

\newcommand{\MM}{{\mathbb M}}
\newcommand{\XX}{{\mathbb X}}

\newcommand{\token}{\mathtt{Tok}}
\newcommand{\ptoken}{\mathtt{p}\token}
\newcommand{\mtoken}{\mathtt{m}\token}

\newcommand{\mpk}{\mathtt{mpk}}
\newcommand{\msk}{\mathtt{msk}}
\newcommand{\bmsk}{\mathtt{bmsk}}
\newcommand{\bmpk}{\mathtt{bmpk}}
\newcommand{\cct}{\mathtt{eRow}}

\newcommand{\ct}{\mathtt{ct}}
\newcommand{\cc}{\mathtt{c}}

\newcommand{\sk}{\mathtt{sk}}

\newcommand{\cgame}{\ensuremath{\mathsf{\aoe Game}}}
\newcommand{\ssdgame}{\ensuremath{\mathsf{SSDGame}}}

\newcommand{\realgame}{{\sf{RealGame}}}
\newcommand{\view}{{\sf{View}}}

\newcommand{\leak}{\mathcal{L}}
\newcommand{\minleak}{\mathsf{mL}}

\newcommand{\simul}{{\sf{Sim}}}

\newcommand{\encrow}{{\mathtt{e}\row}}
\newcommand{\vrow}{{\mathtt{v}\row}}
\newcommand{\vtoken}{\mathtt{v}\token}
\newcommand{\vptoken}{\mathtt{v}\ptoken}
\newcommand{\vmtoken}{\mathtt{v}\mtoken}
\newcommand{\vpol}{\mathtt{v}\pol}

\newcommand{\leakk}{{\mathtt{l}}}       
\newcommand{\lrow}{\leakk\row}
\newcommand{\lacc}{\leakk\mathtt{Req}}
\newcommand{\lpol}{\leakk\pol}
\newcommand{\lk}{\leakk\mathtt{k}}
\newcommand{\lsel}{\leakk\mathtt{Sel}}
\newcommand{\lval}{\leakk\mathtt{Val}}

\newcommand{\metasimul}{{\sf{mSim}}}
\renewcommand{\metasimul}{\simul}

\newcommand{\constr}{{\mathsf{ConstAdm}}}

\newcommand{\ParGen}{{\mathsf{ParGen}}}

\newcommand{\Enc}{{\mathsf{Enc}}}

\newcommand{\Dec}{{\mathsf{Dec}}}
\newcommand{\pDec}{{\mathsf{p}\Dec}}
\newcommand{\mDec}{{\mathsf{m}\Dec}}
\newcommand{\KeyGen}{{\mathsf{KeyGen}}}
\newcommand{\pKeyGen}{{\mathsf{p}\KeyGen}}
\newcommand{\mKeyGen}{{\mathsf{m}\KeyGen}}
\newcommand{\BasicKeyGen}{{\mathsf{B}\KeyGen}}

\newcommand{\symenc}{{\mathsf{enc}}}
\newcommand{\symdec}{{\mathsf{dec}}}
\newcommand{\Init}{{\mathsf{Init}}}
\newcommand{\AuthorizeSel}{{\mathsf{AuthorizeSel}}}
\newcommand{\AuthorizeDec}{{\mathsf{AuthorizeDec}}}
\newcommand{\Encrypt}{{\mathsf{Encrypt}}}
\newcommand{\Select}{{\mathsf{Select}}}
\newcommand{\Decrypt}{{\mathsf{Decrypt}}}

\newcommand{\B}{\mathcal{B}}
\newcommand{\GG}{\mathbb{G}}

\newcommand{\Z}{\mathbb{Z}}

\newcommand{\e}{\mathbf{e}}

\newcommand{\PPol}{\mathbb{P}}
\newcommand{\Conj}{\mathbb{CONJ}}

\newcommand{\innerproduct}[2]{\langle #1,#2\rangle}
\newcommand{\urlGit}{\url{https://github.com/secureselect/SecSel}}

\newcommand{\id}{\mathtt{id}}
\newcommand{\idq}{\id\mathtt{Q}}
\newcommand{\idp}{\id\mathtt{P}}
\newcommand{\ids}{\id\mathtt{S}}
\newcommand{\coal}{\mathsf{C}}
\newcommand{\true}{\mathtt{True}}
\newcommand{\false}{\mathtt{False}}
\newcommand{\flen}{\mathtt{flen}}
\newcommand{\plen}{\mathtt{plen}}
\newcommand{\nlen}{\mathtt{nlen}}

\newcommand{\ctr}{\mathtt{ctr}}

\title{
Secure Selections on Encrypted Multi-Writer Streams
}
\author{
Angelo Massimo Perillo\\ 
Universit\`a di Salerno\\ Salerno\\ Italy\\
\and
Giuseppe Persiano\\
Universit\`a di Salerno\\ Salerno\\ Italy
\and
Alberto Trombetta\\
Universit\`a dell'Insubria\\ Varese\\ Italy
}

\begin{document}
\date{}

\maketitle

\begin{abstract}
Performing searches over encrypted data is a very current and active area. Several
efficient solutions have been provided for the \textit{single-writer} scenario in which 
all sensitive data originates with one party (the \textit{Data Owner}) that encrypts
it and uploads it to a public repository. Subsequently the Data Owner (or authorized
clients, the \textit{Query Sources}) accesses the encrypted data through a \textit{Query Processor} which has direct access to the public encrypted repository.
Motivated by the recent trend in pervasive data collection, we depart from this model and
consider a \textit{multi-writer} scenario in which data originates with several and
mutually untrusted parties. In this new scenario the Data Owner provides public parameters
so that each item of the generated data stream can be put into an encrypted stream;
moreover, the Data Owner keeps some related secret information needed to generate \textit{tokens} so that different \textit{subscribers} can access different subsets
of the encrypted stream in clear, as specified by corresponding access policies.

We propose security model for this problem that we call \textit{Secure Selective Stream}
(\textsf{SSS}) and give a secure construction for it based on hard problems in Pairing-Based Cryptography.
The cryptographic core of our construction is a new primitive, \textit{Amortized Encryption Scheme} (\textsf{AOE}), that is crucial for the efficiency of the resulting \textsf{SSS}.


\end{abstract}
\newpage
\tableofcontents
\newpage

\section{Introduction} 
\label{sec:intro}
As computing devices become more and more pervasive, our means to collect 
data become more and more distributed and allow to have 
data on phenomena that occur in a widespread area. Reality mining is defined
as ``the collection of machine-sensed environmental data pertaining to human
social
behavior~\cite{EaglePentland}'' and has changed the way human interactions are
studied.
A similar phenomenon is taking place in the health care domain in which
epidemiological data can be
collected at a countrywide level by hospitals and private practices.
The ability to create large data sets poses serious privacy concerns and
requires extra care.
Encryption is the obvious tool to preserve data confidentiality and recent
advances in
Cryptography allow the owner of the data to perform 
(or to enable third parties to perform)
specific queries on encrypted data. 
Even though searchable encryption~\cite{DBLP:conf/eurocrypt/BonehCOP04} has
been
introduced as a public key primitive, all systems in the literature 
providing query capabilities over encrypted data (most notably
CryptDB~\cite{DBLP:conf/sosp/PopaRZB11})
have considered a scenario in which the data originates with {\em one} user and
the
same user (or some authorized third parties) performs searches on the encrypted
data.
We are interested in the more challenging scenario in which 
several {\em data sources} generate data managed by a {\em data owner}
and the data owner enables several {\em query sources} to
view parts of the data according to its own access policies (which may vary depending on the query source).
The query sources use {\em query processors} that have direct access to the stream of encrypted data generated by the data
sources and select data to which access is granted.
Thus, roughly speaking, we are interested in a {\em multi-writer}, and thus \emph{
public-key}, setting
whereas previous proposals have considered a {\em single-writer}, and thus \emph{
private-key}, setting.

\begin{figure}[htb]
  \centering
  \includegraphics[scale=0.3]{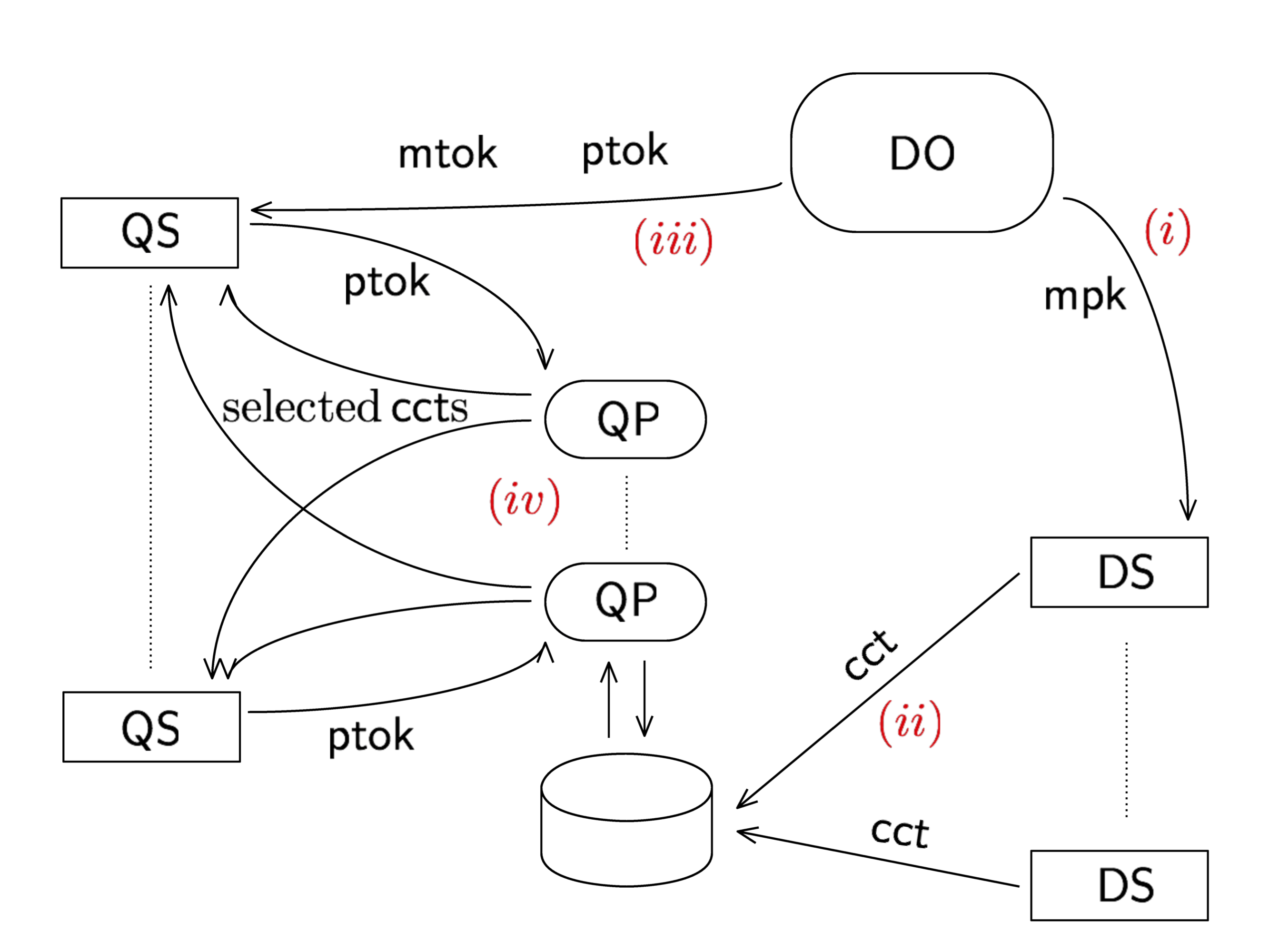}
  \caption{The architecture of our scenario.}
  \label{fig:archi}
\end{figure}

\noindent\emph{Secure Selective Streams.}
We formalize the scenario we just introduced by the notion of a \emph{Secure Selective Stream} (\sss) scheme.
More precisely, we have four different types of actors: 
one \emph{Data Owner}, multiple \emph{Data Sources},
multiple \emph{Query Sources}, multiple \emph{Query Processors} 
(see Figure~\ref{fig:archi}).
The \emph{Data Owner} (the \DO, in short) 
manages access policies to the data originating from several 
\emph{Data Sources} (the \DS s, in short) 
and collected in encrypted form on a possibly untrusted server.
The \emph{Query Processors} (the \QP s, in short) 
have direct physical access to the encrypted data and 
perform the queries on behalf of the \emph{Query Sources} 
(the \QS s, in short).
We consider a threat model in which the Data Owner \DO\ is the only 
fully trusted party. 
The Data Sources (the \DS s) are trusted to upload significant data 
but they should not be able to read the data uploaded by other \DS s.
The Query Processors \QP s are honest-but-curious and 
it is expected to execute the prescribed code.
The Query Sources (the \QS s) should be able to learn only the data 
they have been authorized to read by the \DO.
This requirement extends to coalitions of \QS s: a coalition
can only learn the union of the data they are authorized to read and 
nothing else. Of course, with the help of the \QP s, they could learn, 
for example, which data items were selected by both queries they have been authorized
to issue but, still, no extra data item is revealed.
We also protect the \QS s from the \QP s by not letting the \QP s read the 
result of the queries issued by the \QS s and we want the \QP s not to learn
the exact number of selected data items. 
In other words, the \QP s and \DS\ only learn data-access and search patterns
and no explicit data, except the authorized data, is disclosed.
The mechanism by which the \DO\ decides which query is a 
\QS\ authorized to issue is not considered in this paper. 
We stress also that, even though the \DS s can encrypt data, 
they do not have the ability to authorize searches. 
In other words the ability to write (to encrypt) data is decoupled from the
ability to query (to decrypt) data thus making our scenario inherently
a public-key one.
We look at the case in which the data streams 
are collected as 
\emph{data items} with same number of \emph{cells}.
We aim to support access that correspond to conjunctive 
queries composed by equality-based predicates. 
That is, each query asks to see the content of some of the cells 
of the data items that satisfy the search predicates.

\noindent\emph{Our approach.}
The recent advances in {\em Functional Encryption}~\cite{BonehSW12} provide a 
straightforward secure implementation of our scenario. 
More precisely, the \DO\ publishes the public key of a Functional Encryption
scheme
to be used by a \DS s to encrypt the data items. 
The \DO\ uses the associated secret key to compute the token needed to perform
the query
the specific \QS\ is authorized to perform. 
A \QP\ then simply applies the token to the encrypted data and returns the
result.
This approach has the advantage of supporting any query that can be expressed
by a small (polynomial) circuit~\cite{GargGH0SW13, Waters15, AnanthBSV15} (and,
actually, even more~\cite{AnanthS16}).
Unfortunately, these are to be seen more as feasibility results 
and unlikely to be, at this stage, of direct use in a practical system. 
%
We are less ambitious with respect to the range of queries supported
but we do insist on an efficient and practical solution with clear and provable 
security guarantees.  Specifically, we consider the case in which the 
\DS s generate a stream of {\em rows} with the same number of {\em cells} and we
wish to support queries that select one specified cell from all rows that satisfy an
{\em access policy} that can be expressed as a conjunction of equalities between cells
and constants.
Even for the set of access policies (or, equivalently, queries) of our interest, the state of the art
in public-key functional encryption does not offer an adequate solution.

\noindent
{\em Hidden Vector Encryption} ({\em HVE}, see ~\cite{BW07,IP08,DIP12})
seems to perfectly suit our setting.
Roughly speaking, HVE allows to encrypt plaintext $M$ with 
respect to attribute vector $X=(X_1,\ldots, X_n)$ with components 
taken from an attribute space $\cal X$. 
The owner of the master secret key can generate tokens 
associated with vectors $Y=(Y_1,\ldots,Y_n)$ in which each component
is either a ``don't care symbol'' $\star$ or an element of $\cal X$. 
A token associated with $Y$ can be used to decrypt all ciphertexts 
whose attribute vector $X$ coincides with $Y$ in all components that 
are not $\star$. 
HVE can be used to implement our scenario in a straightforward way:
the \DS\ encrypts each cell of the table by using the values in the other 
columns of the same row as attributes and the value in the cell itself as  
plaintext.
Then, as it easily seen, every query that we wish to support directly maps to a 
vector $Y$ and thus a \QS\ requests the appropriate token to the \DO.
A \QP\ applies the token to each encrypted cell and returns the ones that are 
decrypted correctly.
The simple implementation described above is not practical, though.
First of all, the secret key of all the known implementations of HVE 
need $O(n\cdot\log |{\cal X}|)$ group elements each of size proportional to the security
parameter. 
More importantly, the ciphertext of one cell has length proportional to 
$n\cdot \log |{\cal X}|$ where $n$ is the number of columns in a row.
This implies that a row with $n$ columns, once encrypted, 
will have length $\Omega(n^2)$, clearly impractical.
This second problem seems inherent since, obviously, a ciphertext must be at
	least as long as its attributes.
Our main technical contribution is based on the observation that cells of the same row are 
encrypted using the same attributes and thus we could hope to have an 
{\em amortized} encryption scheme that can be used to reduce the 
cumulative length of the ciphertexts of the cells of a row.

\noindent\emph{Amortized Orthogonality Encryption.}
In order to get around the efficiency drawback of HVE, we introduce the notion 
of an {\em Amortized Orthogonality Encryption} scheme
(an \aoe\ scheme, in short) and provide a construction for it.
Before explaining how $\aoe$ can be used to reach our goal of an efficient and secure $\sss$, we
give a rough description of $\aoe$.
In a regular (i.e., non-amortized) Orthogonality Encryption scheme, ciphertexts and tokens
are associated with attribute vectors of the same length over some finite field.
A token associated with vector $Y$ can be used to decrypt all ciphertexts 
whose associated attribute vector $X$ is orthogonal to $Y$. 
Orthogonality Encryption can be used to implement HVE as well as disjunctive 
queries that will be useful in our scenario.\footnote{Orthogonality Encryption
has been introduced by~\cite{KatzSW08}
with the name of ``Inner Product Encryption.'' 
Recently, the term ``Inner Product'' has been also used by~\cite{AbdallaBCP15} 
in connection with a different albeit related concept.
We choose to use the term ``Orthogonality Encryption'' as
it better reflects the nature of the concept we intend to use and 
avoids any misunderstanding as to which of the two concepts, 
the one of \cite{KatzSW08} or the one of \cite{AbdallaBCP15}, we are using.}
An {\em Amortized} Orthoganility Encryption scheme has the extra feature that
the encryption algorithm takes as input 
$n$ plaintexts $M_1,\ldots,M_n$  and $n+1$ attribute vectors
$X_0,X_1,\ldots,X_n$, 
where $X_1,\ldots,X_n$ have the same length $k$ and  $X_0$ a possibly different
length $l$,
and produces a {\em cumulative ciphertext} $\cct$ 
in which plaintext $M_i$ is encrypted with respect to attribute vector
$(X_0,X_i)$ of length
$l+k$ obtained by concatenating $X_0$ and $X_i$.
The cumulative ciphertext has total length $\Theta(l+n\cdot k)$, which  
is asymptotically optimal as it is proportional to the total length of the
attributes.
The owner of the {\em master secret key} $\msk$ for an \aoe\ can release two
types of tokens:
{\em predicate-only} tokens, the  {\sf P-tokens}, and 
     {\em message} tokens, the {\sf M-tokens}. 
A {\sf P-token} is associated with a vector $Y_0$ of length $l$ and 
it can be used to check whether the attribute vector $X_0$ associated with a 
cumulative ciphertext $\cct$ is orthogonal to $Y_0$. 
Notice that no plaintext is obtained by applying a {\sf P-token} to a $\cct$.
An {\sf M-token} instead is associated with a vector $Y$ of length $l+k$ and,
when applied to a cumulative ciphertext, can be used to obtain message $M_i$ 
only if the corresponding attribute vector $(X_0,X_i)$ is orthogonal to $Y$.

\noindent{\em Efficiency.}
As we have stated above, the cumulative ciphertext has total length
$\Theta(l+n\cdot k)$.
When implemented in a bilinear settings 
(like all the known implementations of HVE and Orthogonality or, 
as it is called in the literature, IPE), 
the length of the ciphertext corresponds to the number of group elements. 
In implementing our scenario for data organized in rows with $n$ columns, 
we will use \aoe\ with $k:=2$ and $l:=2n+1$ thus yielding, for each row,
a cumulative ciphertext with $\Theta(n)$ group elements as opposed to
$\Theta(n^2)$ group elements needed by Orthogonality. 
The saving is not only in space but also in the time needed to perform
encryption and decryption as they take time linear in the number of group 
elements.
Therefore, using \aoe\ guarantees that encryption takes time linear in the
number of columns
whereas Orthogonality would use quadratic time (see also Section~\ref{sec:exp}).

\noindent\emph{Implementing \sss\ using \aoe.}
We use \aoe\ to provide a secure implementation of \sss\, 
according to the following steps (refer to Figure \ref{fig:archi}):
(i) The \DO\ generates a pair of public and secret master key $(\mpk,\msk)$ and 
distributes the $\mpk$ to all \DS s. 
(ii) A \DS\ adds a new data item consisting of cells $M_1,\ldots,M_n$
to the encrypted stream by performing the following steps. 
Each $M_i$ is encrypted by using the
public
master key $\mpk$ and a set of attributes that depends 
on the actual values contained in the cells and on the index $i$ of cell $M_i$.
We point out that the resulting values of $l$ and $k$ will be such that 
a cumulative ciphertext of a data item with $n$ cells has length $\Theta(n)$ 
which is asymptotically optimal (see the discussion in Section~\ref{sec:exp}).
(iii) An access policy $\cal Q$ is specified by the column $d$ to be selected 
and by a sequence of pairs attribute and value $(i_j,m_j)_{j=1}^\nu$, 
for some $\nu\leq n$.
The \DO, upon receiving the request for the token for ${\cal Q}$ from a \QS, 
computes a {\sf P-token} $\ptoken$ and 
a {\sf M-token} $\mtoken$.
The {\sf P-token} checks if the attributes derived from the common attributes 
$M_1,\ldots,M_n$ of the cells of a data item satisfy the predicate 
$$\mathbb{PP}_{\cal Q}(M_1,\ldots,M_n):=
\bigwedge_{j=1}^\nu(M_{i_j}=m_j).$$
The {\sf M-token} instead is such that it can be used to decrypt all cells in
the $d$-th position
of data items that satisfy the predicate 
$$\mathbb{PM}_{\cal
Q}(M_1,\ldots,M_n,i):=(i=d)\wedge\bigwedge_{j=1}^\nu(M_{i_j}=m_j).$$
We postpone the discussion on how $\mtoken$ and $\ptoken$ can be computed by
means of an \aoe.
\DO\ passes $\mtoken$ and $\ptoken$ to \QS\ that keeps the $\mtoken$ for
himself and
passes the $\ptoken$ to a \QP. 
(iv) The \QP\ uses $\ptoken$ to select rows from the encrypted table to be
passed to the \QS. 
(v) The \QS\ applies $\mtoken$ to each of the rows received from the \QP.
We remark that \QP\ does not learn which cell the \QS\ is interested in nor its
content.

We have implemented the construction in C/C++ showing the feasibility and
efficiency of our approach.
We present our implementation in Section \ref{sec:exp}.

\noindent{\em Orthogonality supports $\mathbb{PP}$ and $\mathbb{PM}$.}
Let us now briefly explain how orthogonality can be used to implement
encryption that
supports tokens for $\mathbb{PP}$ and $\mathbb{PM}$.
Inner product computation can be used to encrypt data so that one can issue
tokens to check
polynomial identities. Specifically, observe that evaluating a $d$-degree
multivariate polynomial
$P(x_1,\ldots,x_m)$ in a $m$-dimensional point $(r_1,\ldots,r_m)$ corresponds
to
an inner product computation between the vector $Y$ of the coefficients of the
polynomial $P$
and the $O(m^d)$ monomials $r_1^{d_1}\cdot r_2^{d_2}\cdots\cdot r_m^{d_m}$. 
In our case, the arithmetization of predicate $\mathbb{PP}$ gives a polynomial
of degree $2$ with
with $O(n)$ non-zero coefficients ($n$ is the number of cells in a row) whereas
for
$\mathbb{PM}$ we obtain a polynomial of degree $1$ and thus with $O(n)$
non-zero coefficients.
Therefore, we need \aoe\ with $l=O(n)$ and $k=O(1)$ and this keeps the size of
the cumulative
ciphertext corresponding to a row with $n$ cells $O(n)$. 
As we shall see in Section~\ref{sec:exp} the constants hidden by the asymptotic
notation
are very small and the resulting implementation is quite practical.

\smallskip\noindent
{\em Related works.}
All the major commercial RDBMS releases provide functionalities to encrypt the data they store
(see, for example, \cite{OracleSecurityWhitePaper}). 
However, these solutions are based on data-at-rest encryption,
thus limiting the functionalities over encrypted data, that have to be decrypted by the server 
before queries can be processed.
Therefore this type of solutions is not suited for our scenario.
More limited support for secure operations is provided by systems that manage data streams such as pub/sub systems.
The advent of fast networks and cheap online storage has made viable the management of 
encrypted data at application-level. One of the first works to present the paradigm \emph{database-as-a-service} is
\cite{DBLP:conf/sigmod/HacigumusILM02}. 
In \cite{DBLP:journals/pvldb/BajajS11} a database architecture based on a trusted hardware cryptographic module is presented.
In \cite{DBLP:conf/sigmod/HangKD15}, a prototype is presented that 
executes queries over an encrypted relational database in a multiple client setting. Their approach offers protection to non-compromised clients against a passive attacker that has access to all the data of a fraction of clients. 
{\em Symmetric Searchable Encryption} (SSE) provides a way to perform 
keyword searches on encrypted data. 
Here the Data Owner pre-processes and encrypts the data 
so to allow the Query Processor to perform queries efficiently.
The first construction giving sublinear time was presented in~\cite{Curtmola} 
and extended to conjunctive and general Boolean searches by~\cite{Cash2013}.
Both constructions are {\em single-writer} (only the \DO\ can encrypt) and 
{\em single-reader} (only the \DO\ can perform searches).
This was extended to single-writer and {\em multi-reader} 
(i.e., allowing multiple independent \QS s) by \cite{Jarecki}.
Our system can be seen as the first proposal allowing for 
multiple writers (\DS s) and readers (\QS s).
Having a single trusted writer allows for a centralized and 
optimized pre-processing of the data which is not possible in 
our multi-writer scenario.
On the other hand, most of the proposals based on SSE are static in 
the sense that it is very expensive to add new data 
(unless only single-keyword searches are supported~\cite{KamaraCCS12} 
or extra information is leaked~\cite{Cash2013}).
In contrast, in our proposal any Data Source \DS\ can efficiently add 
new data to the table at the cost of an encryption and
without the help of the \DO.
Different advanced cryptographic primitives have been employed in the design and construction
of systems supporting queries over encrypted tables.
CryptDB~\cite{DBLP:conf/sosp/PopaRZB11,PopaThesis} is the prime example of this line of research.
CryptDB is in the single-writer model and leaks statistical information on the whole 
queried column (that is, not limited to the matching rows) and this can be exploited by 
attacks that can reveal significant information such as repeated values in the column
(see~\cite{kamara15,cryptoeprint:2015:979}).
These attacks leverage on CryptDB's use 
of deterministic and property-preserving encryption (PPE) that are instrumental to support
advanced queries. Our proposal does not use any of these cryptographic primitives and thus
these attacks are irrelevant. Moreover, AOE, unlike PPE, does not have any inherent leakage on the 
encrypted plaintext (besides some trivial information such as plaintext length).
The work of \cite{KamaraMoataz16} provided a new construction that does not make use of 
property-preserving encryption and that supports a large class of SQL queries. 
More recently, Boelter et al.~\cite{2016:568,2016:591} have presented 
a single-writer system for querying encrypted data, called Arx, that supports
range queries, in addition to our set of queries. 
The technical core of the system is a construction of a secure scheme 
for performing range queries on a key-value table~\cite{2016:568}.
Mylar \cite{Mylar} is another recently proposed system that uses advanced encryption techniques
to support web applications that store encrypted data on a server, allows for keyword searches 
over it, and the sharing of data encrypted using different keys.
The system relies on multi-key searchable encryption \cite{cryptoeprint:2013:508} 
that allows for the transformation of a keyword search token from one user to another. 
As shown by Grubbs et al. \cite{cryptoeprint:2016:920}, Mylar can be attacked by compromising the server
allowing the attacker to retrieve the plaintext keyword used for a search and discover which documents contain it. Also, the key transformation technique is transitive, and this can create additional security problems.
These attacks are not relevant for our proposal as we rely on different cryptographic primitives.
Regarding systems managing data streams, such as publish-subscribe systems, overviews of the challenges faced by cryptographic solutions are in
\cite{DBLP:conf/spw/Nikander08} and \cite{DBLP:journals/jcs/YuenSM14}.
There is a growing body of literature applying attribute-based and proxy re-encryption based techniques in pub-sub architectures.
See \cite{DBLP:conf/middleware/PalLKHL12}, \cite{DBLP:journals/fgcs/BorceaGPRR17} for significant examples of
pub-sub systems allowing that collect and distributed encrypted data streams.   

In contrast to the works described above, our work is more cryptographic (and less system oriented) 
in nature and proposes a new efficient cryptographic primitive with a direct application to a 
concrete application scenario that has not been implemented by existing systems.
A first version of the \aoe scheme is presented in \cite{perillo-eurosp17}, where it is deployed 
in order to provide secure queries over an encrypted repository.
In the public-key domain, we mention the first proposal of 
Searchable Encryption~\cite{DBLP:conf/eurocrypt/BonehCOP04}
supporting very simple queries then extended to conjunctive
queries in \cite{BW07}.
Our construction of \aoe\ is inspired by the constructions of 
public-key encryption schemes supporting the 
orthogonality predicate, a concept introduced by~\cite{KatzSW08} along with the first secure 
construction based on bilinear groups of composite order. 
A construction based on bilinear groups of prime order is given in~\cite{park} and it constitutes
the starting block of our \aoe.
Constructions of the orthogonality encryption with adaptive security were 
given in \cite{LOSTW10,OkTa}.
The issue of short ciphertexts and keys for the orthogonality encryption scheme was studied 
in \cite{DBLP:journals/dcc/OkamotoT15} that gave a construction with short ciphertexts 
for the orthogonality encryption but no security guarantee was offered for the attributes
(in other words, \cite{DBLP:journals/dcc/OkamotoT15} gives an {\em attribute-based} orthogonality 
encryption scheme). Note that in our setting this is crucial as the attribute
of a cell are the values of the cells in the same row and therefore they must be kept secret.
The problem of query privacy has also been studied. For the specific case of 
orthogonality encryption, an elegant construction that guarantees security of 
the query has been given in~\cite{SSW09}. 
In general, indistinguishability-based query privacy 
is possible only in a private-key settings
and thus cannot be achieved in our multi-writer scenario;
alternatively, one has to consider the case in which 
the function is sampled from  
a sufficiently large space~\cite{functionPrivate}.

\smallskip\noindent
\paragraph{Roadmap.}
In Section \ref{sec:sss_def}, we give formal definition for the notion of a \sssName.
We also give two security notions and prove equality of the notions for access policies of our interest.

In Section \ref{sec:constr}, we introduce the notion of an \aoeName, and show how to constuct
a \sssName\ using an \aoeName\ as a black box.

In Section \ref{sec:aoeCons}, we give a construction of an \aoeName\ and prove its 
security under hardness assumptions in Bilinear settings.

Finally, in Section \ref{sec:exp} we describe our implementation and discuss the outcome of
our experimental evaluation.

\section{\sssName }
\label{sec:sss_def}
In this section,
we introduce the notion of a
{\em \sssName} (\sss) {\em scheme} and present two
security definitions for it.
We will prove that for a large class of policies,
including the one for which we will provide a construction, the two
security notions coincide.

\subsection{Syntax}
As described in Section \ref{sec:intro}, our scenario for $\sss$ consists of four classes of parties:
one {\em Data Owner} (\DO),
several independent {\em Query Processors} (\QP),
several independent {\em Data Sources} (\DS),
and several independent {\em Query Sources} (\QS).
Data is in the form of {\em rows} with the same number $n$ of {\em cells}.
The Data Owner \DO\ enables access to the data by providing tokens to the Query Sources $\QS$s
who issue access requests consisting of pairs $(\pol,k)$, where $\pol$ is a policy taken from a fixed set of
{\em supported} policies $\PPol$ and $1\leq k\leq n$ is an integer.
Typically, a policy $\pol$ is a predicate evaluated on rows. Access request $(\pol,k)$ asks for access to cell $k$ of all rows $\row$ such that $\pol(\row)=\true$.

\begin{definition}
A {\em \sssName} (\sss) {\em scheme} consists of 6 efficient algorithms
$(\Init,\allowbreak\AuthorizeSel,\allowbreak\AuthorizeDec,
  \allowbreak\Encrypt,\allowbreak\Select,\allowbreak\Decrypt)$
that are used by the parties of our scenario in the following way.
\begin{itemize}
\item The \DO, on input the security parameter $\lambda$
and the length $n$ of the rows, runs
$(\mpk,\msk)\from\Init(1^\lambda,1^n)$  to obtain
the {\em master public key} $\mpk$
and
the {\em master secret key} $\msk$.
The master public key $\mpk$ is given to all Data Sources, whereas the master secret key $\msk$ is kept secret by the \DO.

\item A \DS runs $\cct\from\Encrypt(\mpk,\row)$
to produce an {\em encrypted row} $\cct$ to be placed on the {\em encrypted stream}
that is accessed by the \QP s.

\item Upon receiving an {\em access request} $(\pol,k)$ from a \QS, the \DO\ computes a pair consisting of {\em predicate token} $\ptoken\from\AuthorizeSel(\msk,\pol)$
and {\em message token} $\mtoken\from\AuthorizeDec(\msk,\pol,k)$.
$\ptoken$ can be used to select rows that satisfy the policy $\pol$, whereas $\mtoken$ can be applied to decrypt the $k$-th component of a row that satisfy $\pol$.
The pair $(\ptoken,\mtoken)$ is given to the $\QS$ that has made request.
We expect that $\DO$ checks that the specific \QS\
has the right to request a token for $(\pol,k)$. We do not elaborate further
on this point.

\item The $\QS$ gives the predicate token to a $\QP$ that will use it
to select the rows that satisfy $\pol$ by running
$\{0,1\}\from\Select(\cct,\ptoken)$ on the encrypted rows
that appear in the stream.
The selected rows are passed to the $\QS$.

\item The $\QS$ decrypts the $k$-th component of an encrypted row $\cct$
by running $\row_k\from\Decrypt(\cct,\mtoken,k)$.
\end{itemize}
\end{definition}

We remark that in our model a $\QS$ does not directly access the
encrypted stream but rather it delegates a $\QP$ to select the rows
of interest for the $\QS$ to decrypt. The $\QP$ is not necessarily
trusted and thus will not have access to the stream of data in plain,
not even to the cells of the selected rows that the $\QS$ is authorized
to read.
We stress that our model and implementation are flexible enough
to allow a $\QS$ that has direct access to the encrypted stream
to subsume the role of a $\QP$.

We next give two security definitions for \sss: a simulation-based one and a game-based one.
We shall prove that, for a class of supported policies that we call {\em invertible}, the two notions are equivalent.


\subsection{Simulation-based security}
We start by defining the concept of an {\em instance} of $\sss$ and of a {\em view} of an adversary with respect to an instance.

\begin{definition}
An {\em $(n,m,l)$-instance} $\I=(\stream,\accReq)$ of an $\sss$  with supported set $\PPol$ of policies consists of  two components:
\begin{itemize}
\item
a stream $\stream=((\row_1,\ids_1),\ldots,(\row_m,\ids_m))$ of $m$ pairs
each consisting of a row $\row_i$ with $n$ cells and of the identifier $\ids_i$ of the
$\DS$ that has originated the row;
\item a sequence of {\em access requests}
$\accReq=(\accReq_1,\ldots,\accReq_l)$, where each \\
	$\accReq_j=((\pol_j,k_j),\idq_j,\idp_j)$ consists of an access request $(\pol_j,k_j)$ with $\pol_j\in\PPol$,
of the identifier, $\idq_j$, of the $\QS$ that has issued the $j$-th access request and
of the identifier, $\idp_j$, of the $\QP$ that handles the request on behalf of $\idq_j$.
\end{itemize}
\end{definition}

Let $\coal=\coal_S\cup\coal_P\cup\coal_Q$ be a {\em coalition} consisting of
a set $\coal_S$ of $n_S$ \DS s,
a set $\coal_P$ of $n_P$ \QP s,  and
a set $\coal_Q$ of $n_Q$ \QS s.
We next define $\view^\coal(\lambda,\I)$, the {\em view} in the $\realgame$ of a coalition $\coal$
for an $(n,m,l)$-instance $\I=(\stream,\accReq)$ and security parameter $\lambda$.

\begin{definition}
Let $S$ be an $\sss$.
The view with respect to $S$, $\view^\coal_S(\lambda,\I)$, of a coalition $\coal=\coal_S\cup\coal_P\cup\coal_Q$ for a $(n,m,l)$-instance $\I=(\stream,\accReq)$ and
security parameter $\lambda$ is produced by the following $\realgame^\coal_S(\lambda,\I)$
experiment
\begin{enumerate}
\item Set $(\mpk,\msk)\from\Init(1^\lambda,1^n)$.
\item Write $\stream$ as $\stream=((\row_1,\ids_1),\ldots,(\row_m,\ids_m))$.

	For each $(\row_i,\ids_i)$ with $i\in [m]$

    \quad 
    set $\cct_i\from\Encrypt(\mpk,\row_i)$;

    \quad
    if $\ids_i\in\C_S$ then set $\vrow_i=\row_i$ else
                                    set $\vrow_i=\perp$;

  set $\encrow=(\cct_1,\ldots,\cct_m)$ and
  set $\vrow=(\vrow_1,\ldots,\vrow_m)$.

\item Write
$\accReq$ as
$\accReq=(\accReq_1,\ldots,\accReq_l)$.

	For each $j\in [l]$

    \quad
    write $\accReq_j$ as $\accReq_j=((\pol_j,k_j),\idq_j,\idp_j)$;

    \quad
    set $\ptoken_j\from\AuthorizeSel(\msk,\pol_j)$;

    \quad
    set $\mtoken_j\from\AuthorizeDec(\msk,\pol_j,k_j)$;

    \quad 
    if $\idp_j\in\coal_P$ or $\idq_j\in\coal_Q$ then $\vptoken_j=\ptoken_j$ else
     $\vptoken_j=\perp$;

    \quad 
    if $\idq_j\in\coal_Q$ then $\vmtoken_j=\mtoken_j$ else
     $\vmtoken_j=\perp$;

    \quad 
    set $\vtoken_j=(\vptoken_j,\vmtoken_j)$;

    \quad
    if $\idq_j\in\coal_Q$ then $\vpol_j=(\pol_j,k_j)$;

    \quad
    if $\idq_j\not\in\coal_Q$  and $\idp_j\in\coal_P$ then $\vpol_j=(\pol_j,\perp)$;

    \quad
    if $\idq_j\not\in\coal_Q$  and $\idp_j\not\in\coal_P$ then $\vpol_j=(\perp,\perp)$;

set $\vtoken=(\vtoken_1,\ldots,\vtoken_l)$;

set $\vpol=(\vpol_1,\ldots,\vpol_l)$.

\item Output
$\view_S^\coal(\lambda,\I)=(\mpk,\encrow,\vrow,\vtoken,\vpol).$
\end{enumerate}
\end{definition}

\paragraph{Leakage.}
We next define the {\em minimal} leakage $\minleak(\coal,\I)$ which is obtained by a coalition
$\coal=\coal_S\cup\coal_P\cup\coal_Q$ about a $(n,m,l)$-instance $\I=(\stream,\accReq)$ in \sss.
Roughly speaking, the minimal leakage of an instance $\I$ consists of all data that is either originated by members of the coalition or for which the coalition is authorized.
It consists of the following components:
\begin{enumerate}
\item The parameters $n,m$ and $l$ that are, respectively, the number of cells per row, the total number of rows composing the stream and the number of access requests;
\item Write $\stream$ as $\stream=((\row_1,\ids_1),\ldots,(\row_m,\ids_m))$.

	For each $(\row_i,\ids_i)$, with $i\in[m]$, define $\lrow_i$ as follows:

	\qquad
	if $\ids_i\in\coal_S$ then $\lrow_i=\row_i$ else $\lrow_i=\perp$.

	Define $\lrow=(\lrow_1,\ldots,\lrow_m)$.

\item For each access request $\accReq_j=((\pol_j,k_j),\idq_j,\idp_j)$, with $j\in[l]$,
define $\lpol_j$ and $\lk_j$ as follows:

	\qquad
	if $\idp_j\in\coal_P$ or $\idq_j\in\coal_Q$ then $\lpol_j=\pol_j$ else $\lpol_j=\perp$;

	\qquad
	if $\idq_j\in\coal_Q$ then $\lk_j=k_j$ else $\lk_j=\perp$.

\item For each access request $\accReq_j=((\pol_j,k_j),\idq_j,\idp_j)$, with $j\in[l]$,
  and for each $\row_i$, with $i\in[m]$, define $\lsel_{i,j}$ and $\lval_{i,j}$ as follows:

\qquad
if $\idq_j\in\coal_Q$ or $\idp_j\in\coal_P$ then $\lsel_{i,j}=\lpol_j(\row_i)$
else $\lsel_{i,j}=\perp$;

\qquad
if $\idq_j\in\coal_Q$ and $\pol_j(\row_i)=\true$ then $\lval_{i,j}=\row_{i,k_j}$ else $\lval_{i,j}=\perp$;

Define $\lsel_j=(\lsel_{1,j},\ldots,\lsel_{m,j})$ and
       $\lval_j=(\lval_{1,j},\ldots,\lval_{m,j})$.

\item For each access request $\accReq_j$,
define $\lacc_j=(\lpol_j,\lk_j,\lsel_j,\lval_j)$ and
       $\lacc=(\lacc_1,\ldots,\allowbreak\lacc_l)$.

\item Set $\minleak(\coal,\I)=(\lrow,\lacc)$.

\end{enumerate}

We are now ready for our simulation-based security definition.

\begin{definition}
An $\sss$ is {\em simulation-based secure with respect to leakage $\leak$}
if there exists a probabilistic polynomial-time (PPT) simulator
$\simul$ such that, for all coalitions $\coal$ and $n,m,l=poly(\lambda)$,
the families $$\{\view^\coal(\lambda,\I)\}\text{ and } \{\simul(1^\lambda,\leak(\coal,\I))\}$$
are indistinguishable.
\end{definition}

\begin{definition}
\label{sec_ssd_sim}
\label{sec_ssd}
An $\sss$ is {\em simulation-based secure}
if it is simulation-based secure with respect to minimal leakage $\minleak$.
\end{definition}

\subsection{Game-based security}
\label{game_sec}
Now, we give our second, {\em game-based} security definition.
We model security of \sss\ by means of two games,
$\ssdgame^0$ and $\ssdgame^1$,
between a {\em challenger} $\CH$ and a probabilistic polynomial-time {\em adversary} $\ADV$.
The game $\ssdgame_\ADV^\eta(\lambda)$
for $\eta=0,1$, security parameter $\lambda$ and adversary $\ADV$
starts with $\ADV$ outputting two {\em challenge streams} $\stream_0$ and $\stream_1$ and
a coalition $\coal$ of corrupted players.
$\CH$ receives the {challenge streams},
computes $(\mpk,\msk)\from\Init(1^\lambda,1^n)$ and sends the master public key $\mpk$ to $\ADV$.
Moreover, $\CH$ returns $\estream=\Encrypt(\mpk,\stream_\eta)$ to $\ADV$.
The query phase then starts and $\ADV$ can issue $\accReq_j=((\pol_j,k_j),\idq_j,\idp_j)$, for $j=1,\ldots,l=\poly(\lambda)$,
of its choice in order to receive predicate and message tokens $\ptoken$ and $\mtoken$
from the challenger $\CH$ according to whether $\idq_j,\idp_j\in\coal$.
After $\ADV$ has finished issuing its queries, it outputs bit $b$ and we denote
by $p^\eta_\ADV(\lambda)$ the probability that $\ADV$ outputs $q$.

We let $\inst_0=(\stream_0,\accReq)$ and $\inst_1=(\stream_1,\accReq)$, where
$\accReq=(\accReq_1,\ldots,\accReq_l)$.
We say that $\ADV$ is an {\em admissible adversary}
if $\minleak(\coal,\inst_0)=\minleak(\coal,\inst_1)$.

\begin{definition}
\label{sec_ssd_game}
An $\sss$ $S$ is {\em game-based} secure if, for all admissible PPT adversaries $\ADV$
$$\left|p^0_\ADV(\lambda)-p^1_\ADV(\lambda)\right|\leq\negl(\lambda).$$
\end{definition}

\subsection{Invertible policies}
In this section, we define a class of policies, that we call {\em invertible}, and 
we show that, for $\sss$ supporting an invertible set of policies,
the two notions of security of Definition~\ref{sec_ssd_sim} and Definition~\ref{sec_ssd_game}
are equivalent.

We start by defining the concept of a {\em constraint} and of a {\em compatible} set of constraints.
We identify three types of {\em constraints} for a set $\PPol$ of supported policies over rows of length $n$.
\begin{itemize}
\item
{\em Full Positive Constraint}: $\ctr=(\pol,k,\val)$ consisting of
            policy $\pol\in\PPol$,
            integer $1\leq k\leq n$ and
            value $\val$.
A row $\row=\langle\row_1,\ldots,\row_n\rangle$ is {\em admissible} with respect to $\ctr$
        if $\pol(\row)=\true$ and $\row_k=\val$.

\item {\em Positive Constraint}: $\ctr=(\pol,\perp,\perp)$ consisting of policy $\pol\in\PPol$.
A row $\row$ is {\em admissible} with respect to $\ctr$ if $\pol(\row)=\true$.

\item {\em Negative constraint}: $\ctr=(\pol,\perp,\perp)$ consisting of a policy $\pol\in\PPol$.
A row $\row$ is {\em admissible} with respect to $\ctr$ if $\pol(\row)=\false$.
\end{itemize}

\begin{definition}
A {\em constraint set} $\V=(\ctr^\text{f},\ctr^+,\ctr^-)$ consisting of
a set $\ctr^\text{f}$ of full positive constraints,
a set $\ctr^+$ of positive constraints and
a set $\ctr^-$ of negative constraints is {\em compatible}
if there exists at least one row $\row$ that is admissible with respect to all constraints of $\V$.
\end{definition}

We can now give the following definition of {\em invertible} policies.

\begin{definition}
A set of policies $\PPol$ is {\em invertible} if there exists a polynomial-time algorithm $\constr$ such that,
for all compatible sets of constraints $\V=(\V^\text{f},\V^+,\V^-)$,
outputs a row $\row$ admissible for $\V$.
\end{definition}

The following two sections will prove equivalence of the game-based security notion and the 
simulation-based security notion for \sssName\ supporting invertible policies.

\begin{theorem}
\label{thm:equivalent}
Let $S$ be an $\sss$ supporting an invertible set of policies.
Then $S$ is game-based secure if and only if $S$ is simulation-based secure.
\end{theorem}

\subsection{Simulation-based security implies game-based security for invertible policies}
In this section, we show that if $S$ is simulation-based secure $\sss$
then it is also game-based secure.
Let $\ADV$ be an admissible adversary for the security game $\ssdgame$
and let $\coal$ be coalition output by $\ADV$ and
$\inst_b=(\stream_b,\accReq)$, for $b=0,1$ be the instances defined by $\ADV$.
Note that the two instances have the same minimal leakage
(that is, $\minleak(\inst_0)=\minleak(\inst_1)$).

Now consider an hybrid game in 
in which the encrypted stream $\estream$ is produced
by running the simulator on input the minimum leakage associated with the coalition $\coal$
(instead of setting $\estream$ equal to an encryption of $\stream_0$ or $\stream_1$).
Note that this is possible because the adversary of the security game is admissible and thus
the two instances have the same minimal leakage with respect to $\coal$.
By the simulation-based security, the view  of $\ADV$ in the hybrid game is indistinguishable
from the view of the $\ADV$ in $\ssdgame$ for both $b=0$ and $b=1$.
Therefore the probabilities
$p^0_\ADV(\lambda)$ and $p^1_\ADV(\lambda)$ that $\ADV$ outputs $1$ in  games
$\ssdgame_\ADV^0(\lambda)$ and
$\ssdgame_\ADV^1(\lambda)$, respectively, differ by a negligible factor.

\subsection{Game-based security implies simulation-based security for invertible policies}
\label{game_impl_simul}
For the reverse implication,
we construct a simulator $\simul$ that, for any \sss\ with invertible set of supported policies $\PPol$,
has access to the procedure $\constr$ that takes as input
    a set $\ctr^+$ of satisfied access requests
and a set $\ctr^-$ of unsatisfied access requests and returns a row that satisfies all
constraints.

The simulator $\simul$ takes as input the security parameter $1^\lambda$,
a coalition $\coal=\coal_S\cup\coal_P\cup\coal_Q$ consisting of $\DS s,\QP s$ and $\QS s$ and
the minimal leakage $\minleak(\coal,\inst)$ for a $(n,m,l)$-instance $\inst$.
In addition $\simul$ has black-box access to the algorithms of a $\sss$ implementation.
Roughly speaking, the main difficulty for $\simul$ lies in producing the ciphertexts of the rows that appear
in the stream. The rows that are produced by a corrupted $\DS$ are given in clear as part of the leakage and
thus they can be just encrypted by $\simul$. For each row $\row_i$ that is produced by a honest $\DS$,
$\simul$ uses the leakage received as input to construct a set of constraints
$\V_i=(\ctr_i^\text{f},\ctr_i^+,\ctr_i^-)$ that $\row_i$ must respect. Then, instead of encrypting the actual $\row_i$ appearing
in the instance $\inst$, $\simul$ encrypts a row that is computed by algorithm $\constr$, whose existence is guaranteed
by the hypothesis that the $\sss$ implementation supports an invertible set of policies.
Note that in this way,
the simulator constructs an instance $\inst^\prime$ that has the same leakage as the original instance $\inst$.
Indistinguishability of the output of $\simul$ from the actual view of the coalition $\coal$ then follows from
the assumed game-based security of the implementation.

Let us now formally describe $\simul(1^\lambda,\coal,\minleak(\coal,\inst))$.

\begin{enumerate}

\item Write $\minleak(\coal,\I)$ as $\minleak(\coal,\I)=(\lrow,\lacc)$

\item Set $(\mpk,\msk)\from\Init(1^\lambda,1^n)$.

\item Write $\lrow$ as $\lrow=(\lrow_1,\ldots,\lrow_m)$.

\item For $i\in[m]$ s.t. $\lrow_i\neq\perp$, set $\row_i=\lrow_i$.

\item For $i\in[m]$ s.t. $\lrow_i=\perp$

      \quad Set $\ctr_i^\text{f}=\ctr_i^+=\ctr_i^-=\emptyset$;

      \quad For $j\in[l]$

      \qquad Write $\lacc_j$ as $\lacc_j=(\lpol_j,\lk_j,\lsel_j,\lval_j)$;

      \qquad If $\lsel_{i,j}=\false$ then $\ctr_i^-=\ctr_i^-\cup\{(\lpol_j,\perp,\perp)\}$'

      \qquad If $\lsel_{i,j}=\true$ and $\lk_{i,j}=\perp$ then $\ctr_i^+=\ctr_i^+\cup\{(\lpol_j,\perp,\perp)\}$;

      \qquad If $\lsel_{i,j}=\true$ and $\lk_{i,j}\ne\perp$ then $\ctr_i^\text{f}=\ctr_i^\text{f}\cup\{(\lpol_j,\lk_{i,j},\lval_{i,j})\}$;

      \quad Set $\row_i\from\constr(\ctr_i^{\text{f}},\ctr_i^+,\ctr_i^-)$;

\item For $i\in[m]$ set $\cct_i\from\Encrypt(\mpk,\row_i)$;

\item Set $\cct=(\cct_1,\ldots,\cct_m)$ and $\vrow=\lrow$.

\item For $j\in[l]$

    \quad Write $\lacc_j$ as $\lacc_j=(\lpol_j,\lk_j,\lsel_j,\lval_j)$;

    \quad if $\lpol_j\neq\perp$ then $\vptoken_j\from\AuthorizeSel(\msk,\lpol_j)$ else $\vptoken_j=\perp$;

    \quad if $\lpol_j\neq\perp$ and $\lk_j\neq\perp$ then $\vmtoken_j\from\AuthorizeDec(\msk,\lpol_j,\lk_j)$ else $\vmtoken_j=\perp$;

    \quad set $\vtoken_j=(\vptoken_j,\vmtoken_j)$;

    \quad if $\lpol_j\neq\perp$ and $\lk_j=\perp$ then $\vpol_j=(\lpol_j,\perp)$;

    \quad if $\lpol_j\neq\perp$ and $\lk_j\neq\perp$ then $\vpol_j=(\lpol_j,\lk_j)$;

    \quad if $\lpol_j=\perp$ and $\lk_j=\perp$ then $\vpol_j=(\perp,\perp)$;

set $\vtoken=(\vtoken_1,\ldots,\vtoken_l)$;

set $\vpol=(\vpol_1,\ldots,\vpol_l)$.

\item Return $(\mpk,\encrow,\vrow,\vtoken,\vpol)$.




















\end{enumerate}

\paragraph{Security proof.}\label{equiv_proof}
For the sake of contradiction, we assume the existences of an adversary
$\ADV$ that distinguishes
between $\{\view^\coal_\realgame(\lambda,\I)\}\text{ and }\{\metasimul(1^\lambda,\leak(\coal,\I))\}$.
We then construct a probabilistic polynomial time adversary $\ADVB$ that breaks game $\ssdgame$ thus reaching
a contradiction.

We consider a series of hybrid games $H_0(\lambda,\I),\ldots,H_n(\lambda,\I)$
that are obtained from $\ssdgame$ executed on instances derived from $\I$.
Specifically, for an instance $\I=(\stream,\accReq)$ and for $i=0,\ldots,n$,
in game $H_i(\lambda,\I)$ the first $i$ rows of the stream are constructed by using $\constr$ on the constraint sets
derived from $\accReq$ and the remaining rows are the same as the ones in $\stream$.
Note that $H_0(\lambda,\I)$ coincides with $\view^\coal_\realgame(\lambda,\I)$ and $H_n(\lambda,\I)$ instead coincides with
$\metasimul(1^\lambda,\minleak(\coal,\I))$. Therefore, if $\ADV$ has a non-negligible advantage in distinguishing
$H_0$ and $H_n$, there must exist $i$ such that $\ADV$ has non-negligible advantage
in distinguishing $H_i$ and $H_{i+1}$. That is, by denoting with $p_\ADV^i(\lambda,I)$
and $p_\ADV^{i+1}(\lambda,\I)$
the probabilities that $\ADV$ outputs $1$ when the input is distributed according to
$H_i(\lambda,\I)$ and to $H_{i+1}(\lambda,\I)$,
we have
\begin{equation}
\label{eq1}
\left|p_\ADV^i(\lambda,\I)-p_\ADV^{i+1}(\lambda,\I)\right|\geq 1/\poly(\lambda),
\end{equation}
for some polynomial $\poly$.

Algorithm $\ADVB$ interacts with the challenger $\CH$ of $\ssdgame$ and prepares two streams:
$\stream_0$ is like in $H_i$, that is the first $i$ rows are constructed by using $\constr$ and the remaining
ones are from $\stream$ as appearing in instance $\I$;
$\stream_1$ is like in $H_{i+1}$, that is the first $i$ rows are constructed by using $\constr$ and the remaining
ones are from $\stream$ as appearing in instance $\I$.
Then $\ADVB$ requests from $\CH$ the tokens as specified in $\accReq$ and constructs the view
consisting of all the messages received from $\CH$ and passes it to $\ADV$ receiving a bit $b$.
Then $\ADVB$ outputs $b$ and stops.

We note that $\ADVB$ is an admissible adversary since the leakages for $\I_0=(\stream_0,\accReq)$ and 
$\I_1=(\stream_1,\accReq)$ coincide. Moreover, if $\ADVB$ in engaged in $\ssdgame^0$ with $\CH$,
then the view constructed is exactly the same as in $H_i$
whereas, when engaged in $\ssdgame^1$, it coincides with $H_{i+1}$. 
Therefore, by denoting with $p^0_\ADVB(\lambda)$ and $p^1_\ADVB(\lambda)$ the probabilities
that $\ADVB$ outputs $1$ in $\ssdgame^0$  and $\ssdgame^1$, respectively, we have
that $p^0_\ADVB(\lambda)=p_\ADV^i(\lambda,\I)$ and $p^1_\ADVB(\lambda)=p_\ADV^{i+1}(\lambda,\I)$.
Therefore by Equation~\ref{eq1} we conclude that $\ADVB$ is a successful adversary, thus reaching contradiction.

\section{Constructing $\sss$ from $\aoe$}
\label{sec:constr}
In this section 
we describe a construction of a $\sss$ scheme where cells are elements in $\Z_p$
and the set of supported policies $\Conj$ contains policies expressed as conjunctions of equality predicates.
Specifically, a policy $(\pol,k)\in\Conj$, consists of $\pol=(\pol_1,\ldots,\pol_n)\in(\Z_p\cup\{\star\})^n$,
and, for a row $\row=(\row_1,\ldots,\row_n)$,  we have $\pol(\row)=\true$ iff,
for all $i=1,\ldots,n$, we have that
$$(\pol_i=\star)\lor(\pol_i=\row_i)$$
where $\star$ is a ``don't care'' symbol.

Our construction uses as a black-box an 
{\em Amortized Orthogonality Encryption} ($\aoe$) scheme and prove that the $\sss$ constructed satisfies
the game-based security notion if the $\aoe$ employed is also game-based secure. 

In Section~\ref{sec:aoeCons} we give a construction
of \aoe\ that can be proved secure under hardness assumptions in the Bilinear setting.
In Section~\ref{sec:invertibleconj}, we prove that
the set $\Conj$ of policies is invertible and thus we can conclude that our construction of \sss\ is also
simulation-based secure, under hardness assumptions in the Bilinear setting.

We start by introducing the concept of an {Amortized Orthogonality Encryption} and its security notions.

\subsection{Amortized Orthogonality Encryption}

An \aoe\ scheme is a generalisation of the orthogonality encryption schemes
in which ciphertexts and keys are
associated to {\em attribute} vectors of some fixed length over a finite field.
A key associated with vector $S$ can decrypt a ciphertext associated with vector $X$, iff
$S$ and $X$ are orthogonal. We denote by $\langle X,S\rangle$
the inner product of $X$ and $S$ that checks orthogonality of the two vectors. Also,
for vector $X_0$ of length $n_0$ and vector $X_1$ of length $n_1$, we denote by
$(X_0,X_1)$ the vector of length $n_0+n_1$ obtained by concatenating $X_0$ and $X_1$.
In an {\em Amortized} Orthogonality Encryption scheme (\aoe) the encryption
algorithm takes as input $n$ plaintexts $M_1,\ldots,M_n$ each associated with a vector of length $u+v$ and
the $n$ attribute vectors share the first $u$ components. The goal is to amortize the length of the ciphertexts
so that it is  proportional to $u+n\cdot v$ instead of proportional to $n\cdot (u+v)$.
We will use an \aoe\ with constant $v$ and $u=\Theta(n)$ which will make the total size of the $n$ ciphertext
$\Theta(n)$, a considerable saving over $\Theta(n^2)$.
Let us start by defining the syntax of an $\aoe$ scheme.

\begin{definition}
An \aoe\ scheme with message space $\MM$ and attribute space $\XX$ is a
sequence of $6$ probabilistic polynomial-time algorithms
$(\ParGen,\Enc,\pKeyGen,$ $\mKeyGen,\pDec,\mDec)$ with the following syntax:
\begin{enumerate}
\item the {\em parameter generator} algorithm
$\ParGen(1^\lambda,1^n,1^u,1^v)$ takes as input
{\em security parameter} $\lambda$,
{\em multiplicity factor} $n$, and
{\em length parameters} $u$ and $v$ and
outputs the {\em master public key} $\mpk$ and the {\em master secret key} $\msk$;
%
\item the {\em encryption} algorithm
$\Enc(\mpk,\X,\M)$ takes as input master public key $\mpk$,
a {\em sequence of vectors of attributes} $\X\in{\XX}^u
        \times({\XX}^v)^n$
and a {\em sequence of messages} $\M\in\MM^n$ and outputs {\em cumulative ciphertext}
$\ct=(\ct_0,\ct_1,\ldots,\ct_n)$.
%
\item the {\em {\sf P-token} generator} algorithm $\pKeyGen$ takes as input
the master secret key $\msk$,
and the {\em vector of attributes} $S_0\in\mathbb{X}^u$
and outputs {\sf P-token} $\ptoken$;\\
\item the {\em {\sf M-token} generator} algorithm $\mKeyGen$ takes as input
the master secret key $\msk$,
the {\em vectors of attributes} $S_0\in\mathbb{X}^u$ and
                                    $S_k\in\mathbb{X}^v$ and
integer $k\in\{1,\ldots,n\}$
and outputs {\sf M-token} $\mtoken$;\\
\item the {\em {\sf P}-decryption} algorithm
$\pDec$ takes as input the first component $\ct_0$ of a cumulative ciphertext
$\ct$  and a {\sf P-token} $\ptoken$ and outputs $0$ or $1$;

\item the {\em {\sf M}-decryption} algorithm
$\mDec$ takes as input a pair $(\ct_0,\ct_k)$ of components of a cumulative ciphertext
and an {\sf M-token} $\mtoken$ and outputs either a message $M\in\MM$ or $\perp$.
\end{enumerate}

We have the following two correctness requirements.

\medskip
\noindent\emph{Algorithm $\pDec$:}
For every attribute vector $S_0\in\XX^u$
and
for every sequence $\X=(X_0,X_1,\ldots,X_n)$ of attribute vectors such that
$\langle S_0,X_0\rangle=0$
and for every sequence $\M=(M_1,\ldots,M_n)$ of messages
we have that if
$(\mpk,\msk)\from\ParGen(1^\lambda,1^n,1^u,1^v)$,  and
$\ct\from\Enc(\mpk,\X,\M)$, and
$\ptoken\from\pKeyGen(\msk,S_0)$, then $\pDec(\ct_0,\ptoken)=1$,
except with probability negligible in $\lambda$.

\smallskip
\noindent\emph{Algorithm $\mDec$:}
For every $k\in\{1,\ldots,n\}$, for every attribute vectors
$S_0\in\mathbb{X}^u$ and $S_k\in\mathbb{X}^v$,
for every sequence $\X=(X_0,X_1,\ldots,X_n)$ of attribute vectors
and for every sequence $\M=(M_1,\ldots,M_n)$ of messages
we have that if
$(\mpk,\msk)\from\ParGen(1^\lambda,1^n,1^u,1^v)$,
$\ct\from\Enc(\mpk,\X,\M)$,
and
$\mtoken\from\mKeyGen(\msk,\allowbreak(S_0,S_k,k))$ then,
{if} $\langle(S_0,S_k),(X_0,X_k)\rangle=0$ then
$\mDec((\ct_0,\ct_k),\mtoken)=M_k$, except with probability negligible in $\lambda$.
\end{definition}

\subsection{Security game for \aoe}
We model privacy of the attributes and of the plaintexts in a cumulative ciphertext of an \aoe\
by means of game, \cgame, between a {\em challenger} \CH\ and a probabilistic
polynomial-time {\em adversary} $\ADV$.
The game $\cgame^\ADV(\lambda,n,u,v)$ takes as input the security parameter $\lambda$,
the multiplicity factor $n$ and the length parameters $u$ and $v$.
\begin{enumerate}
\item The game starts with $\ADV$ outputting two {\em challenge} sequences
of attributes ${\cal X}^0=(X_0^0,\ldots,X_n^0)$ and ${\cal X}^1=(X_0^1,\ldots,X_n^1)$,
where $X_0^0$ and $X_0^1$ are attribute vectors of length $u$ and
$X_1^0\ldots,X_n^0,X_1^1\ldots,X_n^1$ have length $v$.

\item \CH\ generates a pair $(\mpk,\msk)$ by running $\ParGen$ on input
$(1^\lambda,1^n,1^u,1^v)$, sends $\mpk$ to $\ADV$
and starts answering $\ADV$ queries.
\item 
$\ADV$ can issue two types of queries that are answered by \CH\ by using $\msk$:
{\sf{M-token} queries} in which
$\ADV$ asks to see the token corresponding to $(S_0,S_k,k)$ of his choice;
and {\sf{P-token} queries} in which $\ADV$ asks to see the token corresponding to
attribute vector $S_0$ of his choice.

\item 
At any time,
$\ADV$ may send two sequences of $n$ messages, $\M^0$ and $\M^1$, to $\CH$ that replies
by flipping a random bit $\xi$ and computing the {\em challenge cumulative ciphertext} $\ct^\star$ corresponding to plaintexts $\M^\xi$ encrypted with attributes $\X^\xi$.

\item 
At the end $\ADV$ outputs its guess $\xi'$ for $\xi$.
\end{enumerate}

We say that $\ADV$ wins the game if $\xi=\xi'$ and 
\begin{enumerate}
\item for all $(S_0,S_k,k)$ for which an {\sf M-token} query has been issued by $\ADV$,
we have that
$\langle (S_0,S_k),(X_0^0,X_k^0)\rangle\ne 0$ {and}
$\langle (S_0,S_k),(X_0^1,X_k^1)\rangle\ne 0$;
\item
for all vectors $S_0$ for which a {\sf P-token} query has been issued by $\ADV$,
we have that
$\langle S_0,X_0^0\rangle=\langle S_0,X_0^1\rangle$.
\end{enumerate}
We denote by  $p^\ADV(\lambda,n,u,v)$ the probability that $\ADV$ wins the game and give the following definition

\begin{definition}
\label{def:aoesec}
An \aoe\ scheme is secure if, for all adversaries $\ADV$ and values $n,u,v=poly(\lambda)$,
$$\left|p^\ADV(\lambda,n,u,v)-\frac{1}{2}\right|$$
is a negligible function of $\lambda$.
\end{definition}
We remark that the notion of security above corresponds to selective attribute hiding and adaptive payload hiding.
\subsection{Constructing \sss}
Now, we construct our
$$\sss=(\sss.\Init,\sss.\AuthorizeSel,\sss.\AuthorizeDec,\sss.\Encrypt,\sss.\Select,\allowbreak\sss.\Decrypt)$$
based on \aoe\ with messages from $\GG_T$ and attributes from $\Z_p$, and
on a symmetric key encryption scheme $\sym=(\symenc,\symdec)$ that takes secret keys and messages
from $\Z_p$, for some prime $p$.

\begin{enumerate}
\item {$\sss.\Init(1^\lambda,1^n)$ algorithm.}
With $\lambda$ and $n$ in input, the algorithm sets $u=n+1$ and $v=2$
and returns $(\mpk,\msk)\from\aoe.\ParGen(1^\lambda,1^n,1^u,1^v)$.

\item {$\sss.\AuthorizeSel(\msk,\pol)$ algorithm.}
The algorithm for generating a selection token takes as input the master secret key $\msk$
and a policy $\pol=(\pol_1,\ldots,\pol_n)$, where, for $i=1,\ldots,n$, $\pol_i\in\Z_p\cup\{\star\}$.
The algorithm,
for $i=1,\ldots,n$, sets $t_i=0$ if $\pol_i=\star$, and otherwise it sets $t_i$ to a random value of $\Z_p$.
The algorithm then constructs the vector of length $u=n+1$
$$S_0=(-t_1,\ldots,-t_n,\sum_{i=1}^n{t_i\pol_i})$$
and computes  and returns $\ptoken\from\aoe.\pKeyGen(\msk,S_0)$.

\item {$\sss.\AuthorizeDec(\msk,\pol,k)$ algorithm.}
The algorithm for generating a decryption token takes as input the master secret key $\msk$, a policy $\pol=(\pol_1, \ldots,\pol_n)$ and integer $k\in\{1,\ldots,n\}$.
A vector $S_0$ of length $u$ is created as for algorithm $\AuthorizeSel$, together with an additional vector $S_k=(k,-1)$ of length $v$. The algorithm returns $\mtoken\from\aoe.\mKeyGen(\msk,S_0,S_k,k)$.

\item {$\sss.\Encrypt(\mpk,\row)$ algorithm.}
The algorithm takes as input the master public key $\mpk$ and a $\row$ with $n$ cells $\row_1,\ldots,\row_n$.
The algorithm randomly selects $\M=M_i\in\GG_T$,
for $i=1,\ldots,n$, and it generates $\sk_i=\HH(M_i)$.
Values $\sk_1,\ldots,\sk_n$ are keys used to encrypt every cell
as $\cc_i\from\sym.\symenc(\sk_i,\row_i)$.
Then, it
sets
$$X_0=(\row_1,\ldots,\row_n,1)$$
and
$$X_i=(1,i)\text{,\quad for }i=1,\ldots,n.$$
Vector $X_0$ has length $u$, whereas the others have length $v$, and all of the vectors compose the sequence of vectors $\X\in\Z_p^u\times(\Z_p^v)^n$, used as input of algorithm $\aoe.\Enc(\mpk,\X,\M)$.
The result $\ct=(\ct_0,\ct_1,\ldots,\ct_n)$ is used together with $\cc=(\cc_1,\ldots,\cc_n)$ to produce the output $\cct=(\ct,\cc)$.

\item {$\sss.\Select(\cct,\ptoken)$ algorithm.}
It takes as input the ciphertexts $\cct=(\ct,\cc)$, where $\ct=(\ct_0,\ct_1,\ldots,\ct_n)$, and a selection token $\ptoken$. It gives as a result the output of $\aoe.\pDec(\ct_0,\ptoken)$.

\item {$\sss.\Decrypt(\cct,\mtoken,k)$ algorithm.}
It takes in input the ciphertexts $\cct$ composed of $\ct=(\ct_0,\ldots,\ct_n)$ and $\cc=(\cc_1,\ldots,\cc_n)$,
a decryption token $\mtoken$ and the integer $k\in\{1,\ldots,n\}$. The algorithm firstly computes the value $M_k\from\aoe.\mDec(ct_0,\ct_k,\mtoken)$. It then retrieves the decryption key $\sk_k$ for the symmetric key encryption scheme by making use of the hash function $\HH(M_k)$. Finally, the message is decrypted and given as output $\row_k\from\sym.\symdec(\sk_k,\cc_k)$.
\end{enumerate}

\subsection{Security of the \sss\ construction}
The proposed construction makes black-box use of the two cryptographic primitives Amortized Orthogonality Encryption $\aoe$
and Symmetric Key Encryption $\sym$.  We next show that if $\aoe$ is game-based secure and $\sym$ is IND-CPA secure
then the construction of $\sss$ is game-based secure.

We proceed by contradiction and assume the existence of an admissible adversary $\ADVA$ that wins the security game of $\sss$
(see Section~\ref{game_sec}) and describe an adversary $\ADVB$ that wins the security game of \aoe.
Our description gives details on how the process is initialized, how $\ADVB$ constructs the challenge ciphertext and
it answers $\ADVA$'s queries for tokens.

We here make use of an \aoe\ with attribute and message spaces $\Z_p$, and of a \sym\ with key and message spaces $\Z_p$.

\medskip\noindent
{\em Init.} $\ADVB$ receives two streams, $\stream^0$ and $\stream^1$, of length $m$,
and coalition $\coal$ of corrupted players from $\ADVA$.
We assume that $\stream^0$ and $\stream^1$ differ in exactly one position. This is without loss of generality
since, if this is not the case, we can consider $m+1$ intermediate streams $\stream_0,\stream_1,\ldots,\stream_m$ where
$\stream_i$ has the first $i$ components equal to the first $i$ components of $\stream^1$ and the remaining $m-i$ equal
to the last $m-i$ components of $\stream^0$. Clearly, $\stream_0=\stream^0$ and $\stream_m=\stream^1$ and,
if $\ADVB$ has a non-negligible advantage in distinguishing $\stream_0$ and $\stream_m$, there
must exist $i$ such that $\ADVB$ has a non-negligible advantage in distinguishing $\stream_{i}$ and $\stream_{i+1}$.
Then observe that $\stream_{i}$ and $\stream_{i+1}$ differ in exactly one component.

We also observe that neither of the two $\DS$s associated to the two components of the two streams is corrupted, otherwise the leakage associated with the two streams would not be equal, thus contradicting the fact
that $\ADVA$ is an admissible adversary. Since neither is corrupted we can assume that they are the same $\DS$.

Finally, without loss of generality, we can assume that the two streams differ in the first component. Thus we can summarize
and write the two streams as

\begin{eqnarray*}
	\stream^0&=&((\row^0_1,\ids_1),(\row^0_2,\ids_2),\ldots,
	(\row^0_m,\ids_m))
\end{eqnarray*}

\noindent and

\begin{eqnarray*}
	\stream^1&=&((\row^1_1,\ids_1),(\row^1_2,\ids_2),\ldots,
	(\row^1_m,\ids_m)).
\end{eqnarray*}

\smallskip\noindent{\em Computing the challenge.}
$\ADVB$ sets $u=n+1$ and $v=2$ and starts game $\cgame(\lambda,n,u,v)$ with $\CH$.
$\ADVB$ starts by computing the two challenge sequences of attributes $\X^0$ and $\X^1$ as done by algorithm
$\Encrypt$ on input $\row^0_1$ and $\row^1_1$, respectively. Specifically, $\ADVB$ sets
$$X_0^0=(\row^0_{1,1},\ldots,\row^0_{1,n},1)$$
and
$$X_0^1=(\row^1_{1,1},\ldots,\row^1_{1,n},1)$$
and $X_i^0=X_i^1=(1,i)$ for $i=1,\ldots,n$.

Moreover, $\ADVB$ randomly selects messages $M_1,\ldots,M_n\in\Z_p$ and sets
$\M^0=\M^1=(M_1,\ldots,M_n)$.

Then $\ADVB$ sends attribute sequences $\X^0=(X_0^0,X_1^0,\ldots,X_n^0)$ and
                                       $\X^1=(X_0^1,X_1^1,\ldots,X_n^1)$ and
message sequences $\M^0$ and $\M^1$ to $\CH$ receiving cumulative ciphertext $\ct_1$ and master public key $\mpk$.

Then, for $i=1,\ldots,n$, $\ADVB$ computes $\cc_{1,i}$ as $\cc_{1,i}=\sym.\symenc(M_i,0)$.
The pair $(\ct_1,\cc_1)$ where $\cc_1=(\cc_{1,1},\ldots,\cc_{1,n})$ constitutes the simulated encryption of the first element
of the stream.  $\ADVB$ then computes the encryption of all the remaining rows by executing the encryption algorithm
$\Encrypt$ using public key $\mpk$.

All the ciphertexts obtained $(\ct_1,\cc_1),\cdots,(\ct_m,\cc_m)$ are then sent to $\ADVA$.

\smallskip\noindent{\em Answering queries.}
Whenever $\ADVA$ issues an access request $\accReq=((\pol,k),\idq,\idp)$,
$\ADVB$ proceeds as follows.

    If $\idq\in\coal$ then $\ADVB$ issues an $\mKeyGen$ request to $\CH$ for $(\pol,k)$ constructing
    a vector of attributes of length $u+v=n+3$ as done by algorithm $\AuthorizeDec$.
    The $\mtoken$ obtained from $\CH$ is then passed to $\ADVA$.
    If $\idq\in\coal$ or $\idp\in\coal$ then $\ADVB$ issues a $\pKeyGen$ request to $\CH$ for $\pol$ constructing
    a vector of attributes of length $u=n+1$ as done by algorithm $\AuthorizeSel$.
    The $\ptoken$ obtained from $\CH$ is then passed to $\ADVA$.

\medskip
The following remarks are in order.
First of all, we observe that $\ADVB$ is an admissible adversary for \aoe\ as all tokens it requests to $\CH$ give
the same result independently from whether $\CH$ has encrypted using the $\X^0$ or $\X^1$.
Indeed $\ADVB$ asks only for tokens that are seen by corrupted players and, by the admissibility of $\ADVA$, they provide
the same results when applied to the two challenges.
Also, observe that the view of $\ADVA$ as constructed by $\ADVB$ is not the same as in the security game for \sss.
Indeed the symmetric ciphertexts corresponding to the row of the first component of the stream are encryptions
of $0$ (and not of the elements of $\row^0_1$ or $\row^1_1$ as in the real view). However, we observe that
the two views are indistinguishable by the IND-CPA security of $\sym$. Further details are omitted.

We thus have the following theorem.
\begin{theorem}
\label{thm:sssgs}
If \aoe\ is game-based secure and \sym\ is IND-CPA secure then $\sss$ is game-based secure.
\end{theorem}

\subsection{$\Conj$ is invertible}
\label{sec:invertibleconj}
In this section we show that the set of policies $\Conj$ is invertible by providing an implementation of the 
algorithm $\constr$. This, together with Theorem~\ref{thm:sssgs} above and Theorem~\ref{thm:equivalent},
gives the following theorem.
\begin{theorem}
\label{thm:sssss}
If \aoe\ is game-based secure and \sym\ is IND-CPA secure then $\sss$ is simulation-based secure.
\end{theorem}

Algorithm $\constr$
takes as input a set of full positive constraints $\ctr^{\text{f}}=(\ctr^{\text{f}}_1,\ldots,\ctr^{\text{f}}_{\flen})$, a set of positive constraints $\ctr^+=(\ctr^+_1,\ldots,\ctr^+_{\plen})$, and a set of negative constraints $\ctr^-=(\ctr^-_1,\ldots,\ctr^-_{\nlen})$. Every policy $\pol\in(\ctr^+\cup\ctr^-)$ is composed of $n$ elements $\pol_1,\ldots,\pol_n\in\Z_p$, where each element is a string for the equality comparison or a don't care symbol $\star$.
The goal of the algorithm is to build a row $\row=(\row_1,\ldots,\row_n)$ that is admissible with respect to all constraints of $\ctr^{\text{f}}\cup\ctr^+\cup\ctr^-$.

Firstly, $\constr$ instantiates an empty row $\row=(\row_1,\ldots,\row_n$) and, for all pairs $k,\val\in\ctr^{\text{f}}$, sets $\row_k=\val$. Then, it solves a system of $\plen$ linear equations where each equation is of the form
$$\pol_1\row_1+\ldots+\pol_n\row_n-\sum_{i=1}^n{\pol_i}=0,$$
with $\pol\in\ctr^+$. We note that the system has at a least one solution, because we know that exists a row complying with those constraints. Now, if $n<=\plen$, the algorithm returns $\row$. Otherwise, we have $d=n-\plen$ cells in the system that can be freely set and are only dependent on the other cells. The procedure picks these cells at random in $\{1,2^\lambda\}$. Then, it checks if all the following inequalities
$$\pol_1\row_1+\ldots+\pol_n\row_n-\sum_{i=1}^n{\pol_i}\neq0,$$
where $\pol\in\ctr^-$, are satisfied. If not, it picks the random values again, until the inequalities are satisfied. We note that only with negligible $d/2^\lambda$ probability the inequalities are not satisfied and the algorithm needs to pick random values again. Finally, the procedure $\constr$ returns $\row$.

\section{Constructing $\aoe$}
\label{sec:aoeCons}
In this section we describe an \aoe\ scheme with
attribute from $\Z_p$, for some large $p$,
and messages from $\GG_T$ and prove that,
under hardness assumptions in bilinear setting, 
the construction satisfies the security property of Definition \ref{def:aoesec}.

We start by formally describing the 6 algorithms defining an $\aoe$.

\begin{enumerate}
\item {$\ParGen(1^\lambda,1^n,1^u,1^v)$ algorithm.}
The algorithm starts by randomly selecting a bilinear mapping
$(\GG,\GG_T,p,\e)$ with security parameter $\lambda$.
Then it randomly selects
$\omega,\alpha_1,\alpha_2,\beta_1,\beta_2\in\Z_p$ and
constructs $n+1$ pairs of {\em basic master secret keys}
$((\bmsk^1_0,\bmsk^2_0),\ldots,(\bmsk^1_n,\bmsk^2_n))$.
The two basic master secret keys in the first pair consist
of $u+1$ quadruples
$
\bmsk^b_0=(\gamma_{i,b,0},\delta_{i,b,0},\theta_{i,b,0},\omega_{i,b,0})_{i=1}^{u+1}
$, for $b=1,2$
and, the reamining $n$ pairs, for $j=1,\ldots,n$ and $b=1,2$
$
\bmsk^b_j=(\gamma_{i,b,j},\delta_{i,b,j},\theta_{i,b,j},\omega_{i,b,j})_{i=1}^{v+1}
$ have length $v+1$.
All components are randomly selected in $\Z_p$ subject to
$\alpha_1\cdot\theta_{i,1,j}-\beta_1\cdot\omega_{i,1,j}=\omega$
and
$\alpha_2\cdot\theta_{i,2,j}-\beta_2\cdot\omega_{i,2,j}=\omega$.
The {\em master secret key} $\msk$ is then set equal to
$$\msk=\Bigl(
    (\alpha_1,\beta_1),(\alpha_2,\beta_2),
    (\omega,g,g_2),(\bmsk_j^1,\bmsk_j^2)_{j=0}^n
        \Bigr),$$
where $g,g_2$ are randomly selected in $\GG$ and $\Lambda=\e(g,g_2)$.
The {\em basic master public keys} $\bmpk^1_0$ and $\bmpk^2_0$ are
\begin{eqnarray*}
\bmpk^1_0&=&(\Gamma_{i,1,0}=g^{\gamma_{i,1,0}},\Delta_{i,1,0} =g^{\delta_{i,1,0}},
          \Theta_{i,1,0}=g^{\theta_{i,1,0}},\\
          &&W_{i,1,0}=g^{\omega_{i,1,0}})_{i=1}^{u+1}
\end{eqnarray*}
\begin{eqnarray*}
\bmpk^2_0&=&(\Gamma_{i,2,0}=g^{\gamma_{i,2,0}},\Delta_{i,2,0}=g^{\delta_{i,2,0}},
           \Theta_{i,2,0}=g^{\theta_{i,2,0}},\\
           &&W_{i,2,0}=g^{\omega_{i,2,0}})_{i=1}^{u+1}
\end{eqnarray*}
\noindent and, for $j=1,\ldots,n$,
\begin{eqnarray*}
\bmpk^1_j&=&(\Gamma_{i,1,j}=g^{\gamma_{i,1,j}},\Delta_{i,1,j}=g^{\delta_{i,1,j}},
           \Theta_{i,1,j}=g^{\theta_{i,1,j}},\\
           &&W_{i,1,j}=g^{\omega_{i,1,j}})_{i=1}^{v+1}
\end{eqnarray*}
\begin{eqnarray*}
\bmpk^2_j&=&(\Gamma_{i,2,j}=g^{\gamma_{i,2,j}},\Delta_{i,2,j}=g^{\delta_{i,2,j}},
           \Theta_{i,2,j}=g^{\theta_{i,2,j}},\\
           &&W_{i,2,j}=g^{\omega_{i,2,j}})_{i=1}^{v+1}.
\end{eqnarray*}

The {\em master public key} $\mpk$ is then computed by setting $\Omega=g^\omega$ and then
seting 
$$\mpk=\Bigl(\Lambda,\Omega,\left(\bmpk_j^1,\bmpk_j^2\right)_{j=0}^n \Bigr).$$

\item {$\Enc(\mpk,\X,\M)$ algorithm.}
The encryption algorithm takes as input the master public key $\mpk$,
a sequence of vectors of attributes $\X=(X_0,\ldots,X_n)$ and a
sequence $\M=(M_1,\ldots,M_n)$ of messages.
Vector $X_0$ has length $u$, whereas the others have length $v$.
The encryption algorithm produces one
{\em cumulative ciphertext} $\ct$ consisting of
$n+1$ {\em basic ciphertexts} $\ct_0,\ldots,\ct_n$.
The encryption algorithm starts by selecting random $y,z_1,z_2\in\Z_p.$
The value $s$ is used to extend the vectors of attributes;
specifically, the algorithm considers the vectors $(X_0,y)$ and $(y,X_j)$ for $j>0$.
We denote by $x_{i,0}$ the $i$-th component of $(X_0,y)$ and,
similarly, by  $x_{i,j}$ the $i$-th component of $(y,X_j)$.
The algorithm also extends the sequence of messages by setting
$M_0=1_{\GG_T}$.
Basic ciphertext $\ct_j$, for $j=0,\ldots,n$,
is computed by first randomly selecting
$l_j,q_j\in\Z_p$ and setting
$$A_j=g^{q_j}\qquad B_j=\Omega^{l_j}\qquad C_j=\Lambda^{q_j}\cdot M_j$$
and then by setting
\begin{eqnarray*}
    D_{i,1,j} &=& W_{i,1,j}^{l_j}\cdot
              \Gamma_{i,1,j}^{q_j}\cdot
             g^{z_1\cdot\alpha_1\cdot x_{i,j}} \\
    E_{i,1,j} &=& \Theta_{i,1,j}^{l_j}\cdot
              \Delta_{i,1,j}^{q_j}\cdot
              g^{z_1\cdot\beta_1\cdot x_{i,j}} \\
    D_{i,2,j} &=& W_{i,2,j}^{l_j}\cdot
              \Gamma_{i,2,j}^{q_j}\cdot
             g^{z_2\cdot\alpha_2\cdot x_{i,j}} \\
    E_{i,2,j} &=& \Theta_{i,2,j}^{l_j}\cdot
              \Delta_{i,2,j}^{q_j}\cdot
              g^{z_2\cdot\beta_2\cdot x_{i,j}}
\end{eqnarray*}
where $i$ goes from $1$ to $u+1$, for $j=0$; and to $v+1$ for $j>0$.

\item {$\mKeyGen(\msk,S_0,S_k,k)$ algorithm.}
The algorithm for generating a message token takes as input
the master secret key $\msk$,
a vector of attributes $S_0$ of length $u$,
a vector of attributes $S_k$ of length $v$, and
integer $k\in\{1,\ldots,n\}$.
The algorithm uses the basic master keys
for $j=0$ and $j=k$ to generate the message token for $(S_0,S_k,k)$.
Instead of considering them separately,
it is useful to see
$\bmsk_0^1$ and $\bmsk_k^1$
as one basic master key of length $f:=(u+1)+(v+1)$ obtained by concatenation
and denote its $i$-th component
$(\gamma_{i,1},\delta_{i,1},\theta_{i,1},\omega_{i,1})$.
Similarly,
$\bmsk_0^2$ and $\bmsk_k^2$  yield a basic master secret key of
length $f$
whose $i$-th component is the quadruple
$(\gamma_{i,2},\delta_{i,2},\theta_{i,2},\omega_{i,2})$.
The algorithm constructs vector of attributes
$S=(S_0,1,-1,S_k)$ of length $f$ and we denote by $s_i$ its $i$-th component.
The algorithm starts by selecting random $\lambda_1,\lambda_2\in\Z_p$.
Then, for $i=1,\ldots,f$,
the algorithm randomly selects $r_{i,1},r_{i,2}\in\Z_p$ and sets
\begin{eqnarray*}
K_{i,1} = g^{\beta_1\cdot r_{i,1}}\cdot g^{\lambda_1\cdot\theta_{i,1}\cdot s_i}
& & L_{i,1} = g^{\alpha_1\cdot r_{i,1}}\cdot g^{\lambda_1\cdot\omega_{i,1}\cdot s_i}\\
K_{i,2} = g^{\beta_2\cdot r_{i,2}}\cdot g^{\lambda_2\cdot\theta_{i,2}\cdot s_i}
& & L_{i,2} = g^{\alpha_2\cdot r_{i,2}}\cdot g^{\lambda_2\cdot\omega_{i,2}\cdot s_i}
\end{eqnarray*}
The algorithm returns message token
$\mtoken=(F,H,(K_{i,1},L_{i,1},K_{i,2},L_{i,2})_{i=1}^f)$, where
$$F=g_2\cdot\prod_{i=1}^f
    K_{i,1}^{-\gamma_{i,1}}\cdot
    L_{i,1}^{-\delta_{i,1}}\cdot
    K_{i,2}^{-\gamma_{i,2}}\cdot
    L_{i,2}^{-\delta_{i,2}}$$
and
   $$H=g^{\sum_{i=1}^f r_{i,1}+r_{i,2}}.$$

\item {$\pKeyGen(\msk,S)$ algorithm.}
The algorithm for generating a predicate token takes as input
the master secret key $\msk$ and
a vector of attributes $S$ of length $u$.
Vector $S$ is extended to length $u+1$ by appending $0$.
The algorithm starts by selecting random $\lambda_1,\lambda_2\in\Z_p$.
Then, for $i=1,\ldots,u+1$,
the algorithm randomly selects $r_{i,1},r_{i,2}\in\Z_p$ and sets
\begin{eqnarray*}
K_{i,1} = g^{\beta_1\cdot r_{i,1}}\cdot g^{\lambda_1\cdot\theta_{i,0,1}\cdot s_i}
& & L_{i,1} = g^{\alpha_1\cdot r_{i,1}}\cdot g^{\lambda_1\cdot\omega_{i,0,1}\cdot s_i}\\
K_{i,2} = g^{\beta_2\cdot r_{i,2}}\cdot g^{\lambda_2\cdot\theta_{i,0,2}\cdot s_i}
& & L_{i,2} = g^{\alpha_2\cdot r_{i,2}}\cdot g^{\lambda_2\cdot\omega_{i,0,2}\cdot s_i}
\end{eqnarray*}
The algorithm returns predicate token\\
$\ptoken=(F,H,(K_{i,1},L_{i,1},K_{i,2},L_{i,2})_{i=1}^{u+1})$, where
$$
F=g_2\cdot\prod_{i=1}^{u+1}
    K_{i,1}^{-\gamma_{i,1}}\cdot
    L_{i,1}^{-\delta_{i,1}}\cdot
    K_{i,2}^{-\gamma_{i,2}}\cdot
    L_{i,2}^{-\delta_{i,2}}$$
and
$$H=g^{\sum_{i=1}^{u+1} r_{i,1}+r_{i,2}}.$$

\item {Decryption algorithms.}
Algorithm $\pDec$ takes as input ciphertext
$\ct_0=(A_0,B_0,C_0,\allowbreak (D_{i,1},E_{i,1},D_{i,2},E_{i,2})_{i=1}^{u+1})$
and $\ptoken=(F,H,(K_{i,1},L_{i,1},K_{i,2},L_{i,2})_{i=1}^{u+1})$
and consists in testing whether the following
product is equal to $1_{\GG_T}$

\begin{eqnarray*}
&&C_0\cdot\e(A_0,F)\cdot \e(B_0,H)\cdot\\
&&\quad\prod_{i=1}^{u+1}\Bigl[
\e(D_{i,1},K_{i,1})\e(E_{i,1},L_{i,1})
\e(D_{i,2},K_{i,2})\e(E_{i,2},L_{i,2}) \Bigr].
\end{eqnarray*}

Algorithm $\mDec$ takes
as input $\ct_0,\ct_k$
and an {\sf{M-token}} $\mtoken$ for $k$.
By setting $f:=(u+1)+(v+1)$
we can write $\mtoken$ as
$$(F,H,(K_{i,1},L_{i,1},K_{i,2},L_{i,2})_{i=1}^f)$$
and, with a slight abuse of notation, we can write
$$\ct=(A_0,B_0,C_0,A_k,B_k,C_k,(D_{i,1},E_{i,1},D_{i,2},E_{i,2})_{i=1}^f).$$

The decryption algorithm returns $M$ computed as
\begin{eqnarray*}
&&M=C\cdot\e(A,F)\cdot\e(B,H)\cdot\\
&&\quad\prod_{i=1}^f
\Bigl[
\e(D_{i,1},K_{i,1})\cdot \e(E_{i,1},L_{i,1})\cdot\\
&&\quad\quad\e(D_{i,2},K_{i,2})\cdot \e(E_{i,2},L_{i,2}) \Bigr]
\end{eqnarray*}
where
$C=C_0\cdot C_k$, $A=A_0\cdot A_k$ and $B=B_0\cdot B_k.$
\end{enumerate}

\subsection{Correctness of the construction}
We show the correctness of the decryption algorithm for {\sf P-token} in the following.
We denote by $\bmsk_{0,1}$ and $\bmsk_{0,2}$ the two basic master secret keys used to
compute both $\ptoken$ and $\ct_0$ and we denote
the $i$-th element of
$\bmsk_{0,1}$ by $(\gamma_{i,1},\delta_{i,1},\theta_{i,1},\omega_{i,1})$ and
the $i$-th element of
$\bmsk_{0,2}$ by $(\gamma_{i,2},\delta_{i,2},\theta_{i,2},\omega_{i,2})$.
We observe that, for $i=1,\ldots,u+1$, we have
\begin{eqnarray*}
\e(D_{i,1},K_{i,1})&\cdot&\e(E_{i,1},L_{i,1})=\\
    \e(g^{l_0\omega_{i,1}},K_{i,1})& \cdot &
    \e(g^{q_0\gamma_{i,1}},K_{i,1}) \cdot
    \e(g^{z_1\alpha_1 x_i},K_{i,1})\\
    \e(g^{l_0\theta_{i,1}},L_{i,1})&\cdot &
    \e(g^{q_0\delta_{i,1}},L_{i,1}) \cdot
    \e(g^{z_1\beta_1 x_i},L_{i,1})         \\
\end{eqnarray*}
and simple manipulations show that the product above is 
$$
    \e(g,g)^{_{\omega l_0r_{i,1}}}
    \e(g,g)^{_{z_1\lambda_1x_is_i\omega}}
    \e(g^{_{q_0\gamma_{i,1}}},K_{i,1})
    \e(g^{_{q_0\delta_{i,1}}},L_{i,1}).
$$
Similarly, we have
\begin{eqnarray*}
&&\e(D_{i,2},K_{i,2})\cdot\e(E_{i,2},L_{i,2})=
    \e(g,g)^{l_0r_{i,2}\omega}\cdot\\
    &&\quad\e(g,g)^{z_2\lambda_2x_is_i\omega} \cdot
    \e(g^{q_0\cdot\gamma_{i,2}},K_{i,2})\cdot
    \e(g^{q_0\cdot\delta_{i,2}},L_{i,2})
\end{eqnarray*}
and therefore
$$
\prod_{i=1}^{u+1} \Bigl[
\e(D_{i,1},K_{i,1})\cdot \e(E_{i,1},L_{i,1})\cdot
\e(D_{i,2},K_{i,2})\cdot \e(E_{i,2},L_{i,2}) \Bigr]$$
is equal to
\begin{eqnarray*}
\e(g,g)^{l_0\omega\sum_{i=1}^{u+1} (r_{i,1}+r_{i,2})}&\cdot&
	\e(g,g)^{_{\omega(z_1\lambda_1+z_2\lambda_2)\cdot\sum_{i=1}^l x_is_i}} \cdot \\
     \prod_{i=1}^{u+1}\Bigl[
                     \e(g^{q_0\cdot\gamma_{i,1}},K_{i,1})& \cdot & \e(g^{q_0\cdot\gamma_{i,2}},K_{i,2}) \cdot \\
                      \e(g^{q_0\cdot\delta_{i,1}},L_{i,1}) &\cdot &
                      \e(g^{q_0\cdot\delta_{i,2}},L_{i,2}) \Bigr].
\end{eqnarray*}
On the other hand, we have
\begin{eqnarray*}
\e(A_0,F)&=&
    \e(g,g_2)^{q_0}\cdot \\
\prod_{i=1}^{u+1}
\Bigl[
      & & \e(g^{-q_0\cdot\gamma_{i,1}},K_{i,1}) \cdot \e(g^{-q_0\cdot\gamma_{i,2}},K_{i,2}) \cdot \\
      & & \e(g^{-q_0\cdot\delta_{i,1}},L_{i,1}) \cdot \e(g^{-q_0\cdot\delta_{i,2}},L_{i,2})
\Bigr]\\
\e(B_0,H)&=&
    \e(g,g)^{-l_0\omega\sum_{i=1}^{u+1}(r_{i,1}+r_{i,2})}.
\end{eqnarray*}
Therefore if $\langle X,S\rangle=0$ we have that the
above product is equal to $\e(g,g_2)^{q_0}=C_0^{-1}$.
On the other hand,
if $\langle X,S\rangle\ne 0$ then the above product is a random element
of $\GG_T$.

\noindent
The decryption algorithm for an {\sf M-token}
is similar to the one for the {\sf P-token} and can be thus omitted.

\subsection{Security of the construction}
In this section, we show that the \aoe\ construction guarantees the security of
the attributes and of the plaintexts of a cumulative ciphertext.

Observe that the encryption algorithm $\Enc$ can be seen as
computing a pair of {\em basic} ciphertexts:
one consisting of the $D_{i,1,j}$ and $E_{i,1,j}$
and
one consisting of the $D_{i,2,j}$ and $E_{i,2,j}$.
The same sequence of vectors of attributes is used for each basic ciphertext.
In the hybrid games we will consider for the security proof,
this will not be necessarily the case
as, in some cases, we will produce challenge ciphertexts
consisting of two basic ciphertexts computed with respect to
two different sequences of vectors of attributes.
Specifically, if $\X^0=(X_0^0,\ldots,X_n^0)$ and $\X^1=(X_0^1,\ldots,X_n^1)$
are two sequences of vectors of attributes then when we say that
the sequence of messages ${\M}=(M_1,\ldots,M_n)$ is encrypted with respect to $(\X^0,\X^1)$,
we actually mean that the first basic ciphertext is with respect to $\X^0$ and the
second with respect to $\X^1$.

The proof uses hybrid games parameterized by $(\lambda,n,u,v)$ and by
a probabilistic polynomial-time algorithm $\ADV$.
The hybrids differ in the way the challenge ciphertext is computed and
are summarized in the following table.
There $\X^0$ and $\X^1$ are the two attribute sequences of length $n+1$
and
${\M}^0$ and ${\M}^1$ are the two sequences  of $n$ messages
given in output by $\ADV$.
We let $\ZZ$ denote the attribute sequence in which
all attribute vectors are zeros;
that is, ${\ZZ}=(0^u,0^v,\ldots,0^v)$.

$$
\begin{array}{|l||c|c|}\hline
\text{Hybrid} & \text{Plaintext} & \text{Attributes} \\ \hline \hline
\cgame_0^\ADV & {\M}^0       &  ({\X}^0,{\X}^0) \\ \hline
\cgame_1^\ADV & \text{random}    &  ({\X}^0,{\X}^0) \\ \hline
\cgame_2^\ADV & \text{random}    &  ({\X}^0,{\ZZ})   \\ \hline
\cgame_3^\ADV & \text{random}    &  ({\X}^0,{\X}^1) \\ \hline
\cgame_4^\ADV & \text{random}    &  ({\ZZ}  ,{\X}^1) \\ \hline
\cgame_5^\ADV & \text{random}    &  ({\X}^1,{\X}^1) \\ \hline
\cgame_6^\ADV & {\M}^1       &  ({\X}^1,{\X}^1) \\ \hline
\end{array}
$$

We stress that the first game, $\cgame_0^\ADV$, coincides with the security game
$\cgame^\ADV$ in which the challenger $\CH$ sets $\xi=0$
and
the last game, $\cgame_6^\ADV$, coincides with the security game
$\cgame^\ADV$ in which the challenger $\CH$ sets $\xi=1$.

Indistinguishability of $\cgame_0^\ADV$ and $\cgame_1^\ADV$ is proved under the BDDH Assumption.
The proof consists in the construction of a probabilistic polynomial-time simulator $\B_1$ that
interacts with a probabilistic polynomial-time adversary $\ADV$
and receives a sequence $\X$ of vectors of attributes,
a sequence of messages $\M$,
and a challenge $({\mathbb B},g,T_1,T_2,T_3,T)$ for the BDDH Assumption.
Simulator $\B_1$ simulates an interaction with $\ADV$ in which the
challenge ciphertext,
depending on the value $\xi$ hidden in the challenge received,
is an encryption of the messages in sequence ${\M}$ with
attributes vectors $\X$ or an encryption of random messages
with attributes vectors $\X$.
By considering $\B_1$ with ${\X}:={\X}^0$ and ${\M}:={\M}^0$
gives that, under the BDDH Assumption, $\cgame_0^\ADV$ and
$\cgame_1^\ADV$ are indistinguishable.
Instead, for indistinguishability of $\cgame_5^\ADV$ and $\cgame_6^\ADV$
we set ${\X}:={\X}^1$ and ${\M}:={\M}^1.$

The remaining hybrids are proved indistinguishable under the BDL Assumption.
Specifically, we construct a probabilistic polynomial-time simulator
$\B_2$ that interacts with a
probabilistic polynomial-time adversary $\ADV$ and receives a challenge
$({\mathbb B},g,T_1,T_2,T_{13},T_4,T)$ for the BDL Assumption. As additional inputs,
$\B_2$ receives two sequences of attribute vectors $\X$ and $\V$.
$\B_2$, depending on the value $\xi$ hidden by the challenge received,
simulates a game in which a randomly chosen sequence of messages is encrypted either with
respect to sequences of attributes $({\X},{\ZZ})$ or with
respect to sequences of attributes $({\X},{\V})$.
Then $\cgame_1^\ADV$ and $\cgame_2^\ADV$ are proved indistinguishable
by running $\B_2$ on
sequences ${\X}:={\X}^0$ and ${\V}:={\X}^0$.
To prove indistinguishability of
$\cgame_2^\ADV$ and $\cgame_3^\ADV$, $\B_2$ is run on
sequences ${\X}:={\X}^0$ and ${\V}:={\X}^1$.
To prove indistinguishability of
$\cgame_3^\ADV$ and $\cgame_4^\ADV$,
and
$\cgame_4^\ADV$ and $\cgame_5^\ADV$,
we use a mirror image of $\B_2$ (that we call $\B_2'$)
that, depending on the value $\xi$ hidden by the challenge received,
encrypts a randomly chosen sequence of messages
either with respect to sequences of attributes $({\ZZ},{\X})$ or with
respect to sequences of attributes $({\V},{\X})$.
Thus, to prove indistinguishability of
$\cgame_3^\ADV$ and $\cgame_4^\ADV$,
$\B_2'$ is run on
sequences ${\X}:={\X}^1$ and ${\V}:={\X}^0$ and
to prove indistinguishability of
$\cgame_4^\ADV$ and $\cgame_5^\ADV$, $\B_2'$ is run on
sequences ${\X}:={\X}^1$ and ${\V}:={\X}^1$.

\subsubsection{Simulators}
Simulators $\B_1$ and $\B_2$ are described in the following.
The description of $\B_2'$ can be obtained by modifying $\B_2$ in a
straightforward way and is thus omitted.

\paragraph{Simulator $B_1$.}
It takes as input a challenge $(g,T_1=g^{t_1},T_2=g^{t_2},T_3=g^{t_3},T=\e(g,g)^{t_1t_2t_3+\xi\cdot r})$
for the BDDH assumption along with
a sequence $\M=(M_1,\ldots,M_n)$ of messages and
a sequence $\X=(X_0,X_1,\ldots,X_n)$ of vectors of attributes.
$\B_1$ interacts with adversary $\ADV$ and,
depending on whether $\xi=0$ or $\xi=1$, simulates
$\cgame^\ADV(\lambda,n,u,v)$ in which
the challenge ciphertext is an encryption with attributes $\X$ of $n$ random elements of
$\GG_T$ or of the messages in $\M$.

\par\noindent{\em Constructing $\mpk$ and a partial $\msk$.}
The master secret key contains $(n+1)$ pairs of {\em basic} master secret keys $(\bmsk_j^1,\bmsk_j^2)$.
We let $\ell_j$ denote the length of the basic master secret keys of the
$j$-th pair and write
$\bmsk_j^1=
    (\gamma_{i,1,j}, \delta_{i,1,j},\theta_{i,1,j},\omega_{i,1,j})_{i=1}^{\ell_j}$
and
$\bmsk_j^2=
    (\gamma_{i,2,j}, \delta_{i,2,j},\theta_{i,2,j},\omega_{i,2,j})_{i=1}^{\ell_j}$.
Clearly, $\ell_0=u+1$ and $\ell_j=v+1$ for $j>0$.

$\B_1$ starts by randomly selecting
$\rho_0,\ldots,\rho_n\in\Z_p$ and
$\alpha_1,\alpha_2,\beta_1,\beta_2,\omega,y\in\Z_p$.
We let $x_{i,0}$ denote the $i$-th component of the vector $(X_0,y)$ of length $u+1$;
and, for $j>0$, we let $x_{i,j}$ denote the $i$-th component of the vector $(y,X_j)$ of length $v+1$.

For
    $i=1,\ldots,\ell_j$, $b=1,2$, and $j=0,\ldots,n$,
$\B_2$ picks
random $\tilde\gamma_{i,b,j},\tilde\delta_{i,b,j},\theta_{i,b,j},\omega_{i,b,j}\in\Z_p$
subject to
$$\alpha_b\cdot\theta_{i,b,j}-\beta_b\cdot\omega_{i,b,j}=\omega.$$
$\B_1$ computes the basic master secret keys in such way to implicitly set
$$\begin{array}{lclclcl}
\gamma_{i,b,j}=x_{i,j}\cdot\alpha_b\cdot\frac{t_2}{\rho_j}+\tilde\gamma_{i,b,j} \\
\delta_{i,b,j}=x_{i,j}\cdot\beta_b \cdot\frac{t_2}{\rho_j}+\tilde\delta_{i,b,j} \\
\end{array}$$
It is easy to verify that the values are independent and uniformly
distributed in $\Z_p$.
Clearly, the values $\gamma_{i,b,j}$ and $\delta_{i,b,j}$ cannot be explicitly computed
by $\B_1$ but it is easy to see that $\Gamma_{i,b,j}=g^{\gamma_{i,b,j}}$ and
$\Delta_{i,b,j}=g^{\delta_{i,b,j}}$ can be computed by setting
$$\begin{array}{lclclcl}
\Gamma_{i,b,j}=T_2^{x_{i,j}\cdot\alpha_b\cdot\frac{1}{\rho_j}}\cdot g^{\tilde\gamma_{i,b,j}}\\
\Delta_{i,b,j}=T_2^{x_{i,j}\cdot\beta_b \cdot\frac{1}{\rho_j}}\cdot g^{\tilde\delta_{i,b,j}}\\
\end{array}$$
and, obviously, $W_{i,b,j}=g^{\omega_{i,b,j}}$ and $\Theta=g^{\theta_{i,b,j}}$, for all $i,b$ and $j$.
Finally, $\B_1$ sets $\Omega=g^\omega$ and, instead of setting $\Lambda=\e(g,g_2)$,
picks a random $\eta\in\Z_p$ and sets
$\Lambda=\e(T_1,T_2)^\omega\cdot\e(g,g)^\eta$.
The value of $g_2$ is thus implicitly set equal to $g^{\eta+\omega\cdot t_1\cdot t_2}.$


\noindent{\em Answering token queries.}
We next describe how $\B_1$ answers
{\sf M}-token queries for $(S_0,S_k,k)$.
{\sf P}-token queries for vector $S_0$ are simpler and can be handled similarly.
We remind the reader that an {\sf M}-token is computed by constructing
the vector $(S_0,1,-1,S_k)$ of length $f:=(u+1)+(v+1)$ whose
$i$-th component will be denoted by $s_i$.
For the sake of a more agile notation, we will
collapse corresponding values from $\bmsk_0$ and $\bmsk_d$ into
one single vector of length $f$.
For example, instead of considering vectors $(\omega_{i,1,0})_{i=1}^{u+1}$ and
$(\omega_{i,1,d})_{i=1}^{v+1}$ as two separate vectors, we will consider the vector of
length $f$ obtained by concatenating them and denote by $\omega_{i,1}$ its $i$-th component.
Similarly, for $\omega_{i,2},\theta_{i,1},\theta_{i,2},
\tilde\gamma_{i,1},\tilde\gamma_{i,2},\gamma_{i,1},\gamma_{i,2},
\tilde\delta_{i,1},\tilde\delta_{i,2},\delta_{i,1},$ and $\delta_{i,2}$.


$\B_1$ sets $c=2\cdot\left(\innerproduct{X_0}{S_0}/\rho_0+
                           \innerproduct{X_k}{S_k}/\rho_k\right)$,
randomly selects $\tilde\lambda_1,\tilde\lambda_2\in\Z_p$ and
returns a token with the same distribution as the token
returned by $\mKeyGen$ with randomness
$$\lambda_1=\tilde\lambda_1+\frac{t_1}{c}\qquad
  \lambda_2=\tilde\lambda_2+\frac{t_1}{c}.$$
Observe that,
since $\innerproduct{(X_0,X_k)}{(S_0,S_k)}\ne 0$ and
$\rho_0$ and $\rho_k$ are random in $\Z_p$, the probability that $c=0$ is
negligible. Moreover, if $c\ne 0$, $\lambda_1$ and $\lambda_2$ are independent and
uniform in $\Z_p$.

Specifically, for $i=1,\ldots,f$,
$\B_1$ randomly selects $r_{i,1},r_{i,2}\in\Z_p$ and sets
$$\begin{array}{lcl}
K_{i,b}=g^{-\beta_1\cdot r_{i,b}+\tilde\lambda_b\cdot\theta_{i,b}\cdot s_i}\cdot
		T_1^{\frac{\theta_{i,b}\cdot s_i}{c}}\\
L_{i,b}=g^{\alpha_1\cdot r_{i,b}-\tilde\lambda_b\cdot\omega_{i,b}\cdot s_i}\cdot
		T_1^{-\frac{\omega_{i,b}\cdot s_i}{c}}
\end{array}$$
As the values $r_{i,1},r_{i,2}$ are known to $\B_1$ for all $i$,
$H$ can be computed in a straightforward way.
The computation of $F$ requires some more care.
Given $K_{i,1},K_{i,2}$ and $L_{i,1},L_{i,2}$
and $\gamma_{i,1}$ and $\delta_{i,1}$,
$F$ is expected to be equal to
$g_2\cdot\prod_{i=1}^f K_{i,1}^{-\gamma_{i,1}}\cdot L_{i,1}^{-\delta_{i,1}}\cdot
                K_{i,2}^{-\gamma_{i,2}}\cdot L_{i,2}^{-\delta_{i,2}}.$
By setting,
$\mu_i=\rho_0$ for $i=1,\ldots,u+1$ and $\mu_i=\rho_k$ for $i=u+2,\ldots,f$,
and, for $i=1,\ldots,f$ and $b=1,2$,
$$
\begin{array}{lcl}
Z_{i,b}=\omega_{i,b}\cdot\beta_b-\theta_{i,b}\cdot\alpha_b\\
\Phi_{i,b}=\omega_{i,b}\cdot\tilde\delta_{i,b}-\theta_{i,b}\cdot\tilde\gamma_{i,b}\\
\Psi_{i,b}=\beta_b\cdot\tilde\gamma_{i,b}-\alpha_b\cdot\tilde\delta_{i,b}
\end{array}
$$
we can we write
\begin{eqnarray*}
F=&g_2\cdot&\prod_{i=1}^f K_{i,1}^{-\gamma_{i,1}}\cdot L_{i,1}^{-\delta_{i,1}}\cdot
                K_{i,2}^{-\gamma_{i,2}}\cdot L_{i,2}^{-\delta_{i,2}} \\
 =&g_2\cdot&\left[g^{t_1\cdot t_2}\right]^{-2\cdot\omega\sum_i \frac{s_i\cdot x_i}{\mu_i\cdot c}}
      \cdot\left[g^{t_1}\right]^{\sum_{i=1}^f \frac{(\Phi_{i,1}+\Phi_{i,2})\cdot s_i}{c}} \cdot  \\
    & & \left[ g^{t_2}\right]^{
    		\sum_{i=1}^{f}(\tilde\lambda_1\cdot Z_{i,1}+\tilde\lambda_2\cdot Z_{i,2})\cdot x_i\cdot s_i\cdot\frac{1}{\mu_i}}
		\cdot  \\
     & & g^{\sum_{i=1}^f \left[r_{i,1}\cdot\Psi_{i,1}+r_{i,2}\cdot\Psi_{i,2}+s_i\cdot(\tilde\lambda_1\cdot\Phi_{i,1}+\tilde\lambda_2\cdot\Phi{i,2})\right]} \\
\end{eqnarray*}
Also, observe that, by definition of $c$, we have that
$$2\cdot \sum_{i=1}^f\frac{x_i\cdot s_i}{\mu_i}=c$$
and therefore the exponent of $g^{t_1t_2}$ in the expression above is equal to $-\omega.$
Now, by recalling that $g_2$ has been implicitly set equal to $g_2=g^{\eta+\omega\cdot t_1\cdot t_2}$,
$F$ is expected to have value
\begin{eqnarray*}
F&=&g^\eta\cdot
         \left[g^{t_1}\right]^{\sum_{i=1}^f \frac{(\Phi_{i,1}+\Phi_{i,2})\cdot s_i\cdot\mu_i}{2\cdot c_x}} \cdot  \\
   & & \left[ g^{t_2}\right]^{
    		\sum_{i=1}^{f}(\tilde\lambda_1\cdot Z_{i,1}+
                           \tilde\lambda_2\cdot Z_{i,2})\cdot
        \frac{x_i\cdot s_i}{\mu_i}}
		\cdot\\
      & & g^{\sum_{i=1}^f \left[r_{i,1}\cdot\Psi_{i,1}+r_{i,2}\cdot\Psi_{i,2}+s_i\cdot(
        \tilde\lambda_1\cdot\Phi_{i,1}+
        \tilde\lambda_2\cdot\Phi{i,2})\right]}
\end{eqnarray*}
and $\B_1$ has all it is required to compute $F$ according to the expression above.

\noindent{\em Preparing the challenge ciphertexts.}
$\B_1$ randomly picks $\tilde z_1,\tilde z_2\in\Z_p$ and,
$\tilde l_j$, for $j=0,\ldots,n$.
$\B_1$ prepares the challenge ciphertexts as if they were generated
by $\Enc$ with randomness
$z_1=\tilde z_1-t_2\cdot t_3,z_2=\tilde z_2-t_2\cdot t_3$
and, for $j=0,\ldots,n$,
$l_j=\tilde l_j, q_j=\rho_j\cdot t_3.$
This is obtained by setting
$
    D_{i,b,j}=g^{\tilde l_j\cdot\omega_{i,b,j}}\cdot
            T_3^{\tilde\gamma_{i,b,j}\cdot\rho_j}\cdot
            g^{\alpha_b\cdot\tilde z_b\cdot x_{i,j}}$
and
    $E_{i,b,j}=g^{\tilde l_j\cdot\theta_{i,b,j}}\cdot
            T_3^{\tilde\delta_{i,b,j}\cdot\rho_j}\cdot
            g^{\beta_b\cdot\tilde z_b\cdot x_{i,j}}.
$
Indeed, observe that we can write $D_{i,b,j}$ as
\begin{eqnarray*}
D_{i,b,j}&=&
g^{\tilde l_j\cdot\omega_{i,b,j}}\cdot g^{\rho_j\cdot t_3\cdot [\gamma_{i,b,j}-x_{i,j}\cdot\alpha_b\cdot\frac{t_2}{\rho_j}]}
\cdot g^{\tilde z_b\cdot x_{i,j}\cdot\alpha_b}\\
&=&g^{\tilde l_j\cdot\omega_{i,b,j}}\cdot g^{\rho_j\cdot t_3\cdot \gamma_{i,b,j}}\cdot
    g^{[\tilde z_b-t_2\cdot t_3]\cdot x_{i,j}\cdot\alpha_b}
\end{eqnarray*}
and, similarly, $E_{i,b,j}$ can be written as
$$E_{i,b,j}=
g^{\tilde l_j\cdot \theta_{i,b,j}}\cdot g^{\rho_j\cdot t_3\cdot \delta_{i,b,j}}\cdot g^{[\tilde z_b-t_2\cdot t_3]\cdot x_{i,j}\cdot\beta_b}.$$
Finally, for $j=0,\ldots,n$, $\B_2$ sets
$$A_j=T_3^{\rho_j}\qquad B_j=g^{\tilde l_j\omega}$$ and
$$C_j=T^{-\omega\cdot\rho_j}\cdot \e(g,T_3^{\rho_j})^{-\eta}\cdot M_j.$$
Note that the settings above are equivalent to setting $A_j=g^{q_j}$ and $B_j=g^{l_j\cdot\omega}$.
Let us now look at $C_j$.
Clearly, if $\xi=1$, then $T$ is random in $\GG_T$ and the challenge ciphertext constructed by $\B_1$
is the cumulative ciphertext of sequence of random messages.
Let us consider the case in which $\xi=0$ and thus $T=\e(g,g,)^{t_1\cdot t_2\cdot t_3}$.
In this case, $C_j$ is expected to have value
\begin{eqnarray*}
\e(g,g_2)^{-q_j}\cdot M_j & = & e(g,g)^{-(\eta+\omega\cdot t_1\cdot t_2)\cdot t_3\cdot\rho_j}\cdot M_j\\
&   & (\text{since\ } g_2=g^{\eta+\omega\cdot t_1\cdot t_2})\\
& = & T^{-\omega\cdot\rho_j}\cdot \e(g,T_3^{\rho_j})^{-\eta} \cdot M_j
\end{eqnarray*}
which is exactly the value computed by $\B_1$.

\paragraph{Simulator $\B_2$.}
It takes as input a challenge
$(g,T_1=g^{t_1},T_2=g^{t_2},T_{13}=g^{t_1\cdot t_3},T_4=g^{t_4},T=g^{t_2\cdot(t_3+t_4)+\xi\cdot r})$ for the Bilinear Decision Linear assumption
and two sequences of vectors of attributes
${\X}=(X_0,X_1,\ldots,X_n)$,
${\V}=(V_0,V_1,\ldots,V_n)$.
We remind the reader that $X_0$ and $V_0$ have length $u$ and
$X_i$ and $V_i$ have length $v$, for $i=1,\ldots,n$.
Simulator $\B_2$ interacts with adversary $\ADV$ and,
depending on whether $\xi=0$ or $\xi=1$, simulates
$\cgame^\ADV(\lambda,n,u,v)$ in which
the challenge ciphertext is the encryption of $n$ random elements of
$\GG_T$ with  attributes $({\X},{\ZZ})$ or $({\X},{\V})$ .

$\B_2$ starts by randomly selecting
$\rho_0,\ldots,\rho_n\in\Z_p$ and
$\alpha_1,\alpha_2,\beta_1,\beta_2,\tilde\omega,y\in\Z_p$.
We let $x_{i,0}$ denote the $i$-th component of the vector $(X_0,y)$ of length $u+1$;
and, for $j>0$, we let $x_{i,j}$ denote the $i$-th component of the vector $(y,X_j)$ of length $v+1$.
Similarly for $v_{i,j}$.

\noindent{\em Constructing $\mpk$ and a partial $\msk$.}
We next show how $\B_2$ determines the basic master secret keys.
For
    $i=1,\ldots,\ell_j$, $b=1,2$, and $j=0,\ldots,n$,
$\B_2$ picks
random $\tilde\gamma_{i,b,j},\tilde\delta_{i,b,j},\tilde\theta_{i,b,j},\tilde\omega_{i,b,j}\in\Z_p$
subject to
$$\alpha_b\cdot\tilde\theta_{i,b,j}-\beta_b\cdot\tilde\omega_{i,b,j}=\tilde\omega.$$
$\B_2$ computes the basic master secret keys in such way to implicitly set
$\gamma_{i,1,j}=\tilde\gamma_{i,1,j}$ and $\delta_{i,1,j}=\tilde\delta_{i,1,j}$,
$\gamma_{i,2,j}=\alpha_2\cdot v_{i,j}\cdot t_2/\rho_j+\tilde\gamma_{i,2,j}$ and
$\delta_{i,2,j}=\beta_2 \cdot v_{i,j}\cdot t_2/\rho_j+\tilde\delta_{i,2,j}$ and
$$\begin{array}{lclclcl}
\theta_{i,b,j}=\beta_1 \cdot x_{i,j}\cdot t_2/\rho_j+\tilde\theta_{i,1,j}\cdot t_1\\
\omega_{i,b,j}=\alpha_1\cdot x_{i,j}\cdot t_2/\rho_j+\tilde\omega_{i,1,j}\cdot t_1\\
\theta_{i,2,j}=\beta_2 \cdot v_{i,j}\cdot t_2/\rho_j+\tilde\theta_{i,2,j}\cdot t_1\\
\omega_{i,2,j}=\alpha_2\cdot v_{i,j}\cdot t_2/\rho_j+\tilde\omega_{i,2,j}\cdot t_1.
\end{array}$$
It is easy to verify that the exponents are independently and uniformly distributed over $\Z_p$.
Clearly, $\B_2$ can only partially compute the basic master secret keys
(as they involve values from the Decision Linear challenge tuple).
The master public key instead can be computed by the following settings.
$$\begin{array}{llll}
\Omega=T_1^\omega & \qquad \\
\Gamma_{i,1,j}=g^{\gamma_{i,1,j}} &
\Delta_{i,1,j}=g^{\delta_{i,1,j}} \\
\Theta_{i,1,j}=T_2^{\frac{\beta_1\cdot x_i}{\rho_j}} \cdot T_1^{\tilde\theta_{i,1,j}} &
     W_{i,1,j}=T_2^{\frac{\alpha_1\cdot x_i}{\rho_j}}\cdot T_1^{\tilde\omega_{i,1,j}}  \\
\Gamma_{i,2,j}=T_2^{\frac{\alpha_2\cdot v_i}{\rho_j}}\cdot g^{\tilde\gamma_{i,2,j}} &
\Delta_{i,2,j}=T_2^{\frac{ \beta_2\cdot v_i}{\rho_j}}\cdot g^{\tilde\delta_{i,2,j}} \\
\Theta_{i,2,j}=T_2^{\frac{ \beta_2\cdot v_i}{\rho_j}}\cdot T_1^{\tilde\theta_{i,2,j}} &
     W_{i,2,j}=T_2^{\frac{\alpha_2\cdot v_i}{\rho_j}}\cdot T_1^{\tilde\omega_{i,2,j}}.
\end{array}$$
Simple computation shows that the  settings above are
compatible with the implicit settings of the master secret key.
Moreover, we note that, for $j=0,\ldots,n$, $b=1,2$ and $i=1,\ldots,\ell_j$, we have
\begin{eqnarray*}
\alpha_b\theta_{i,b,j}-\beta_b\omega_{i,b,j}&=&
\alpha_b( \beta_b\chi_{i,j} t_2/\rho_j+\tilde\theta_{i,b,j} t_1)-\\
 &&\ \beta_b(\alpha_b\chi_{i,j} t_2/\rho_j+\tilde\omega_{i,b,j} t_1) \\
&=&t_1(\alpha_b\tilde\theta_{i,b,j}-\beta_b\tilde\omega_{i,b,j}) \\
&=&t_1\tilde\omega:=\omega,
\end{eqnarray*}
where $\chi_{i,j}$ is $x_{i,j}$ or $v_{i,j}$, depending on whether $b=1$ or $b=2$.

\noindent{\em Answering token queries.}
Let us now describe how $\B_2$ constructs the replies to {\sf M}-token queries for $(S_0,S_k,k)$.
A {\sf P}-token is constructed in a similar way. We omit further details.
We use the same notation as for simulator $\B_1$.

$\B_2$ starts by computing $c_x,c_w\in\Z_p\setminus\{0\}$ such that
\begin{equation}\label{eq:cxcy}
c_w\cdot\langle (S_0,S_k),(X_0,X_k)\rangle=
  c_x\cdot\langle (S_0,S_k),(W_0,W_k)\rangle.
\end{equation}
Notice that if one of the two inner product is $0$, so is the other and
thus it is always possible to choose $c_x$ and $c_w$.
If both inner products are equal to $0$, $\B_2$ takes $c_x=c_w=1$.
$\B_2$ then picks random $\tilde\lambda_1,\tilde\lambda_2\in\Z_p$ and, for each $i$,
random values $\tilde r_{i,1},\tilde r_{i,2}\in\Z_p$ and constructs
the values $K_{i,1},L_{i,1},K_{i,2},L_{i,2}$ as if they were computed by
$\BasicKeyGen$ with randomness
$\lambda_1=\tilde\lambda_1-\frac{\tilde\lambda_2}{t_1}$,
$\lambda_2=\frac{\tilde\lambda_2}{t_1}$ and, for  $i=1,\ldots,f$,
$$
\begin{array}{ll}
r_{i,1}=\tilde r_{i,1}-c_w\cdot\frac{\tilde\lambda_2\cdot x_i\cdot s_i\cdot \tau_{i}}{t_1}\\
r_{i,2}=\tilde r_{i,2}\cdot t_1+c_x\cdot\frac{\tilde\lambda_2\cdot v_i\cdot s_i\cdot \tau_{i}}{t_1}
\end{array}$$
where $x_i,v_i$ and $s_i$ respectively denote the $i$-th component of vectors
$(X_0,y,y,X_j)$, $(W_0,y,y,W_j)$, and $(S_0,1,-1,S_j)$ and
$\tau_i=t_2/\rho_0$ for $i=1,\ldots,u+1$ and
$\tau_i=t_2/\rho_j$ for $i=u+2,\ldots,f$.
Simple computation shows that
$K_{i,1},L_{i,1},K_{i,2},L_{i,2}$ can be computed in the following way
$$\begin{array}{lcl}
K_{i,1}=T_1^{\tilde\lambda_1\cdot s_i\cdot\tilde\theta_{i,1}}\cdot
        T_{2,i}^{\tilde\lambda_1\cdot s_i\cdot x_i\cdot\beta_1}\cdot
        g^{-\beta_1\cdot \tilde r_{i,1}}\cdot
        g^{-\tilde\lambda_2\cdot s_i\cdot\tilde\theta_{i,1}}\\
L_{i,1}=T_1^{-\tilde\lambda_1\cdot s_i\cdot\tilde\omega_{i,1}}\cdot
        T_{2,i}^{-\tilde\lambda_1\cdot s_i\cdot x_i\cdot\alpha_1}\cdot
        g^{\alpha_1\cdot\tilde r_{i,1}}\cdot
        g^{\tilde\lambda_2\cdot s_i\cdot\tilde\omega_{i,1}} \\
K_{i,2}=T_1^{-\beta_2\cdot\tilde r_{i,2}}\cdot
        g^{\tilde\lambda_2\cdot s_i\cdot\tilde\theta_{i,2}}\\
L_{i,2}=T_1^{\alpha_2\cdot\tilde r_{i,2}}\cdot
        g^{-c_x\cdot\tilde\lambda_2\cdot s_i\cdot\tilde\omega_{i,2}}\\
\end{array}$$
where,
for $i=1,\ldots,u+1$ and for $b=1,2$, we set
$$T_{2,i}=T_2^{1/\rho_0}\qquad\tilde\theta_{i,b}=\theta_{i,b,0}\qquad\tilde\omega_{i,b}=\theta_{i,b,0}$$
for $i=u+2,\ldots,f$ and for $b=1,2$, we set
$$T_{2,i}=T_2^{1/\rho_j}\qquad\tilde\theta_{i,b}=\theta_{i-(u+1),b,j}\qquad\tilde\omega_{i,b}=\theta_{i-(u+1),b,j}.$$

Let us now show how $\B_2$ computes the two remaining values $F$ and $H$.
As far as $H$ is concerned, we observe that each $i$ contributes the factor
$g^{-(\tilde r_{i,1}+\tilde r_{i,2})}$ and that
\begin{eqnarray*}
\tilde r_{i,1}+\tilde r_{i,2}&=&r_{i,1}+t_1r_{i,2}+\frac{\tilde\lambda_2\cdot t_2}{t_1}\cdot
    \left ( c_xs_iv_i-c_ws_ix_i\right)\\
&=&r_{i,1}+t_1 r_{i,2}
\end{eqnarray*}
where we have used Equation~\ref{eq:cxcy}.
Therefore $H$ can be computed by multiplying, for all $i$, the factors $g^{r_{i,1}}\cdot T_1^{r_{i,2}}$.
Let us now concentrate on $F$. Each $i$ contributes to $F$ the factor
$K_{i,1}^{-\gamma_{i,1}}\cdot L_{i,1}^{-\delta_{i,1}}\cdot
 K_{i,2}^{-\gamma_{i,2}}\cdot L_{i,2}^{-\delta_{i,2}}$ and note that
$\gamma_{i,1}$ and
$\delta_{i,1}$ are known to $\B_2$.
We next show that $\B_2$ can compute the remaining component. Indeed, $\B_2$ needs to
compute
\begin{eqnarray*}
 K_{i,2}^{-\gamma_{i,2}}\cdot L_{i,2}^{-\delta_{i,2}}&=&
\left[T_1^{-\beta_2\cdot \tilde r_{i,2}}\cdot g^{\tilde\lambda^2\cdot s_i\cdot\tilde\theta_{i,2}}\right]^{\alpha_2\cdot\chi_i\cdot \tau_2+\tilde\gamma_{i,2}} \\
& & \cdot
\left[T_1^{\alpha_2\cdot \tilde r_{i,2}}\cdot g^{-\tilde\lambda^2\cdot s_i\cdot\tilde\omega_{i,2}}\right]^{\beta_2\cdot\chi_i\cdot \tau_2+\tilde\delta_{i,2}}
\end{eqnarray*}
but notice that the unknown, to $\B_2$, term $T_1^{\tau_2}$ cancels out leaving only terms that can be computed using
quantities available to $\B_2$.
This completes the description of how $\B_2$ answers token queries.

\noindent{\em Preparing the challenge ciphertexts.}
We remind the reader that, for $j=0,\ldots,n$,
the challenge ciphertext consists of $A_j, B_j, C_j$ and a pair
$(\ct_{1,j},\ct_{2,j})$ of basic ciphertexts each consisting of $\ell_j$
pairs of elements. More precisely, we write
$\ct_{1,j}=(D_{i,1,j},E_{i,1,j})_{i=1}^{\ell_j}$
and
$\ct_{2,j}=(D_{i,2,j},E_{i,2,j})_{i=2}^{\ell_j}.$
Let us now describe how the pair of basic ciphertexts is computed.
$\B_2$ randomly picks $\tilde z_1,\mu_0,\ldots,\mu_n\in\Z_p$ and computes
the $j$-th pair of basic ciphertexts as if output
by the $\Enc$ algorithm with randomness
$l_j=t_3\cdot\rho_j$ and $q_j=t_4\cdot\rho_j+\mu_j$ and
$$\tilde z_1=\tilde z^1-t_2\cdot t_3\qquad z^2=r.$$
Notice that $z_1$ and $z_2$ and $l_j$'s and $q_j$'s are independently and uniformly distributed in $\Z_p$.

For each $j$, $\B_2$ sets
$A_j=T_4^{\rho_j}\cdot g^{\mu_j}=g^{t_4\cdot\rho_j+\mu_j}=g^{q_j}$,
$B_j=T_{13}^{\rho_j\cdot\omega}=g^{\tilde\omega\cdot t_3\cdot \rho_j}=g^{l_j\cdot\tilde\omega}$
and randomly selects $C_j$ at random in $\GG_T$.
The components, $D_{i,1,j}$ and $E_{i,1,j}$, of the first basic ciphertext are computed as
\begin{eqnarray*}
D_{i,1,j}
&=& T_{13}^{\rho_j\omega_{i,1,j}} T_4^{\rho_j\gamma_{i,1,j}}
    g^{\alpha_1 x_{i,j} \tilde z_1} g^{\mu_j\gamma_{i,1,j}}\\
	&=& g^{_{t_3\rho_j(\omega_{i,1,j} t_1+t_2\alpha_1 x_{i,j}/\rho_j)}}
	g^{_{(t_4\rho_j+\mu_j)\tilde\gamma_{i,1,j}}} \cdot\\
	& & \qquad\cdot
    g^{\alpha_1 x_{i,j}(\tilde z_1-t_2 t_3)}\\
&=& g^{l_j\omega_{i,1,j}}
    g^{q_j\gamma_{i,1,j}}
    g^{z_1\alpha_1 x_{i,j}}\\
	E_{i,1,j}&=&
T_{13}^{\rho_j\tilde\theta_{i,1,j}}\cdot T_4^{\rho_j\cdot\tilde\delta_{i,1,j}}\cdot
    g^{\beta_1\cdot x_{i,j}\cdot\tilde z_1}\cdot g^{\mu_j\cdot\tilde\delta_{i,1,j}}
\end{eqnarray*}
Simple algebraic manipulations, similar to the ones used for $D_{i,1,j}$, show that
$E_{i,1,j}=
g^{l_j\cdot\theta_{i,1,j}}\cdot g^{q_j\cdot\delta_{i,1,j}}\cdot
   g^{z_1\cdot\beta_1\cdot x_{i,j}}.$
Therefore $D_{i,1,j}$ and $E_{i,1,j}$ are distributed exactly claimed.
The components, $E_{i,2,j}$ and $D_{i,2,j}$,
of the second basic ciphertext are computed as
\begin{eqnarray*}
D_{i,2,j}
&=&T_{13}^{\rho_j\tilde\omega_{i,2,j}} T_4^{\rho_j\tilde\gamma_{i,2,j}}
    T^{\alpha_2 v_{i,j}}
            g^{\mu_j\tilde\gamma_{i,2,j}} g^{\mu_j\alpha_2 v_{i,j} t_2/\rho_j} \\
&=&g^{t_3\rho_j\tilde\omega_{i,2,j} t_1} g^{t_4\rho_j\tilde\gamma_{i,2,j}}
     g^{\alpha_2 v_{i,j} [t_2(t_3+t_4)+r]}\\
& &
      \cdot   g^{\mu_j\tilde\gamma_{i,2,j}} g^{\mu_j\alpha_2 v_{i,j} t_2/\rho_j} \\
&=&g^{t_3\rho_j(\tilde\omega_{i,2,j} t_1+\alpha_2 v_{i,j} t_2/\rho_j)}
   g^{\tilde\gamma_{i,2,j}(t_4\rho_j+\mu_j)}  \\
& & \cdot
   g^{\alpha_2 v_{i,j} t_2/\rho_j t_4\rho_j}
   g^{\alpha_2 v_{i,j} t_2/\rho_j\mu_j} g^{r\alpha_2 v_{i,j}}\\
&=&g^{t_3\rho_j(\omega_{i,2} t_1+\alpha^2\chi_i t_2/\rho_j)}\cdot \\
	& &\qquad\cdot g^{(t_4\rho_j+\mu_j)(\gamma_{i,2}+\alpha^2\chi_i t_2/\rho_j)}
	g^{r\alpha^2 v_{i,j}}\\
&=&g^{l_j\omega_{i,2,j}} g^{q_j\gamma_{i,2,j}} g^{r\alpha_2 v_{i,j}}\\
E_{i,2,j}&=&
   T_{13}^{\rho_j\tilde\theta_{i,2,j}} T_4^{\rho_j\tilde\delta_{i,2,j}}
    g^{\mu_j\beta_2 v_{i,j}t_2/\rho_j+
    \mu_j\tilde\delta_{i,2,j}} T^{\beta^2v_{i,j}}
\end{eqnarray*}
Simple algebraic manipulations give
$$E_{i,2,j}=
g^{l_j\cdot\theta_{i,2,j}}\cdot g^{q_j\cdot\delta_{i,2,j}}\cdot
   g^{r\cdot\beta_2\cdot v_{i,j}}.$$
Finally, observe that if $r=0$ the ciphertext produced corresponds to $n$ random
messages encrypted with attributes $(\X,\ZZ)$. If instead $r$ is random,
then $\B_2$ has produced the cumulative ciphertext corresponding to $n$
random messages encrypted with attributes $(\X,\V)$.

\section{Experimental evaluation}
\label{sec:exp}
Our experimental work consists of a complete implementation in C/C++  of
our \aoe\ and of an implementation of our scenario
(see Figure~\ref{fig:archi}).
We use the MIRACL library\footnote{\url{https://github.com/miracl/MIRACL}}, 
freely provided by CertiVox (now MIRACL). 
Our implementation uses a Barreto-Naehrig curve 
over a 256-bit field~\cite{Barreto2006} that gives a security level equivalent
to AES-128.
The source code is available at {\tt \urlGit}.

We remind the reader that our data is organized as a stream of rows with the same number of cells
and that each row is encrypted independently from the other rows, possibly by
different \DS s.
For a stream composed of rows with $m$ columns, we instantiate an \aoe\ that can encrypt
$n:=m$ plaintexts (one for each cell of the row to be encrypted) 
with $l:=2m+1$ common attributes and $k:=2$ specific attributes. 
Let us see how $l$ and $k$ are determined by describing the encryption procedure
(Step {\em ii} of the scenario). 
Consider a row consisting of $m$ strings $c_1,\ldots,c_m$ (the cells).
For $i=1,\ldots,m$, the encryption procedure randomly selects 
an element $g_i$ of the target group $\GG_T$ and then obtains a 128-bit key
$k_i$ by
hashing the selected element 
(this is done by using the {\tt hash\_to\_aes\_key} function of the MIRACL
library).
Finally, $c_i$ is encrypted with AES in CBC mode using $k_i$ as a key.
The $m$ group elements $g_1,\ldots,g_m$  are then encrypted with our \aoe\ 
(remember that the message space of our \aoe\ implementation coincides with the 
target group $\GG_T$). The attributes used to encrypt the $g_i$'s are
derived from the strings $c_1,\ldots,c_m$ and from the noise $R$.
Remember that the attribute space of our \aoe\ coincides with $\Z_p$ 
($p$ is the prime order of the bilinear setting used) and thus
we first hash each $c_i$ to obtain $x_i\in\Z_p$.
The common attributes used to encrypt the cells are 
obtained from the $m$ integers $x_1,\ldots,x_m$ and from $R$ and they
correspond to the values of the monomials that could possibly have 
a non-zero coefficient in the polynomial resulting from the arithmetization of 
the predicate corresponding to the queries supported by our system 
(the arithmetization is, for example, illustrated in~\cite{KatzSW08}). 
It is straightforward to see that, for every supported query $\cal Q$, 
the polynomial corresponding to predicate $\mathbb{PP}_{\cal Q}$ has at most
$2\cdot m+1$
non-zero terms (i.e., $x_1,\ldots,x_m,R$ and $R\cdot x_1,\ldots,R\cdot x_m$).
For example, 
the query ${\cal Q}$ from the previous example that asks for the {\tt userId} of all
patients with fever and nausea with $R'=2$ corresponds to polynomial
$(R-2)\cdot[r_5\cdot(x_5-1)+r_6\cdot (x_6-1)]$ obtained by arithmetizing predicate
$\mathbb{PP}_{\cal Q}$, for randomly chosen $r_5,r_6$.
This is expanded into a vector of $13=2m+1$ entries with 6 non-zero entries
(corresponding to coefficients of $R,x_5,x_6$, of $R\cdot x_5$, of $R\cdot x_6$ and 
the zero-degree term).
The polynomial corresponding to predicate $\mathbb{PM}_{\cal Q}$ 
contributes two more terms (the one corresponding to the clause $(i=d)$).
Therefore, we have $l=2m+1$ and $k=2$ for a total of $2m+1+2m=4m+1=\Theta(m)$ attributes.
Firstly, we provide experimental evidence that \aoe\ has better 
performance than (non-amortized) Orthogonality encryption and this justifies
the introduction of \aoe\ as a new primitive.
We report below the multiplicative blow-up in time for the encryption function
of Orthogonality with respect to \aoe\ and in space of the Orthogonality
ciphertext with
respect to the \aoe\ ciphertext.

\begin{tabular}{rrr}
{\tt\#cols} &  {\tt time } & {\tt memory} \\
16  &  1.56 &  2.38\\
32  &  2.96 &  4.5 \\
64  &  6.09 &  8.98 \\
128 & 12.11 & 17.79 \\
\end{tabular}

Thus, for rows with $32$ cells $\aoe$ is faster by almost a factor of $3$ 
and uses $4.5$ times less memory. The gap between the two 
implementations widens as the number of cells in a row grows. 
This is expected since, as remarked in the ``Efficiency'' paragraph of Section I.B,
$\aoe$ yields row encryption with linear (in the number of cells)
time and space complexity whereas
(non-amortized) Orthogonality encryption gives quadratic complexity.
We remark that the time needed to encrypt a row does not depend on how similar the cells of
the row are.

The aim of the next experiment is to collect data on the time needed
by each operation and times must not be considered in absolute terms 
but only to have an idea of the time needed to process a cell. 
This is particularly true for encryption as it will be performed 
by \DS s that will likely batch encrypt only a few rows every time.
Significant is the running time of the {\sf P-token} application and this 
is a direct consequence of the fact that the distributed nature 
of our scenario does not allow for the option of pre-processing the 
data before uploading it to the cloud repository. 
We remark also that it is reasonable to 
assume that the \QP\ (the party applying the {\sf P-token})
is the one with the most computing power. Moreover, 
the fact that the {\sf P-token} can be applied to each row independently
makes it amenable of a highly parallel implementation. 
The next table reports an estimate of the time per cell in milliseconds 
for each of the four operations. The time for token generation is the time to
generate both tokens.

\begin{center}
{\small 
\begin{tabular}{|l|r|}
\hline
Operation & ms per cell/column \\\hline\hline
Key Generation & 1\\
Encryption    &   3.7          \\
Token Generation &  51     \\
Applying a {\sf P-token} &     1.82           \\
Applying an {\sf M-token} &   2.5            \\\hline
\end{tabular}
}
\end{center}

\section{Conclusions}
\label{sec:conc}
In this work we have introduce the notion of
a Secure Stream Selection scheme \sss. Here a number of
potentially untrusted parties can encrypt elements in a stream data
that can be selectively decrypted by third parties authorized by a
trusted data owner. 
We give two notions of security, a game-based one and a simulation-based one,
that are shown to be equivalent for a
class of access policies that include policies of interest for applications.

We give a construction of \sss\ on top of Amortized Orthogonality Encryption scheme \aoe\ that permits the efficient encryption of several data items with respect to
sets of attributes that differ in few elements.
We have proved that our construction is secure under standard assumptions in a bilinear setting. We have provided an implementation that shows the feasibility and effectiveness of our approach, using the C/C++ MIRACL library.

\end{document}